\documentclass[
    reprint,
    superscriptaddress,
    amsmath,amssymb,
    aps,
    pre
]{revtex4-2}

\usepackage[utf8]{inputenc}
\usepackage[T1]{fontenc}

\usepackage{graphicx}
\usepackage{tikz}

\usepackage{multirow}
\usepackage{booktabs}
\usepackage{dcolumn}

\usepackage{bm}

\usepackage{enumitem}

\usepackage{hyperref}


\begin{document}

\preprint{APS/123-QED}

\title{Shaping chaos in bilayer graphene cavities}

\author{Jucheng Lin}
\thanks{These authors contributed equally to this work.}
\affiliation{Department of Physics, Harvard University, Cambridge, Massachusetts 02138, USA}
\affiliation{Department of Chemistry and Chemical Biology, Harvard University, Cambridge, Massachusetts 02138, USA}
\affiliation{Department of Physics and Astronomy, Shanghai Jiao Tong University, Shanghai 200240, China}

\author{Yicheng Zhuang}%

\thanks{These authors contributed equally to this work.}

\affiliation{Department of Physics, Harvard University, Cambridge, Massachusetts 02138, USA}
\affiliation{Department of Chemistry and Chemical Biology, Harvard University, Cambridge, Massachusetts 02138, USA}
\affiliation{Department of Physics, School of Physics, Peking University, Beijing 100871, China}

\author{Anton M. Graf}
\affiliation{Department of Physics, Harvard University, Cambridge, Massachusetts 02138, USA}
\affiliation{Department of Chemistry and Chemical Biology, Harvard University, Cambridge, Massachusetts 02138, USA}
\affiliation{Harvard John A. Paulson School of Engineering and Applied Sciences, Harvard University, Cambridge, Massachusetts 02138, USA}

\author{Eric J. Heller}
\email{eheller@fas.harvard.edu}
\affiliation{Department of Physics, Harvard University, Cambridge, Massachusetts 02138, USA}
\affiliation{Department of Chemistry and Chemical Biology, Harvard University, Cambridge, Massachusetts 02138, USA}

\author{Joonas Keski-Rahkonen}
\affiliation{Department of Physics, Harvard University, Cambridge, Massachusetts 02138, USA}
\affiliation{Department of Chemistry and Chemical Biology, Harvard University, Cambridge, Massachusetts 02138, USA}

\date{\today}

\begin{abstract}
Bilayer graphene cavities where electrons are confined within finite graphene flakes provide an alluring platform not only for the future nanoelectronic devices owing to the tunable energy gap but also for investigating the quantum nature of chaos due to the trigonal warping of their Fermi surface. Here we demonstrate that rotating the cavity boundary relative to the underlying lattice structure drives a quantum transition from nearly integrable dynamics to chaotic regime, observed as a concomitant crossover of eigenvalue statistics and eigenstate profiles. Complementing the full quantum treatment, we examine the classical backbone of this onset of chaos by employing semiclassical ray dynamics. Our results position bilayer graphene cavities as a promising venue for investigating and engineering quantum-chaotic behavior in graphene-based devices.
\end{abstract}

\maketitle

\section{Introduction}
Most physical systems are nonintegrable and can exhibit chaotic behavior in their classical limit~\cite{ott2002chaos}. The huge qualitative differences in integrable and chaotic behavior can be pivotal  to the function of quantum devices. Billiard models have long served as canonical model systems for studying such behavior, in which dynamical complexity arises solely from the geometry of the confining boundary~\cite{chernov2006chaotic}.
In the quantum regime, billiards correspond to idealized cavities governed by the Schrödinger equation with boundary conditions encoding information on the underlying classical motion, thereby providing a direct window into the quantum signatures of chaos~\cite{Stockmann_book, Heller_book, Gutzwiller_book, Haake_book, Tabor_book, Casati_book, Nakamura_book}. This paradigm has played an essential role in establishing the semiclassical link between chaotic classical dynamics and quantum features, paving the way for random matrix theory (RMT)~\cite{mehta2004random} as a general organizing principle for chaotic quantum systems. This perspective was further crystallized in the Berry conjecture~\cite{berry1977regular} and in Bohigas–Giannoni–Schmit (BGS) conjecture~\cite{bohigas1984characterization}, facilitated by subsequent works, such as of Berry and Tabor~\cite{berry_proc.r.soc.lond.a_356_375_1997} as well as Casati~\cite{casati1980connection}. Consequently, billiard systems have become a foundational testbed for various quantum-chaotic phenomena~\cite{Stockmann_book, Heller_book, Gutzwiller_book, Haake_book, Tabor_book, Casati_book, Nakamura_book}, ranging from the universality of level statistics and their periodic-orbit corrections to wavefunction ergodicity and its notable exceptions, such as quantum scarring~\cite{kaplan2026paradoxical}.

Nearly all of the aforementioned results are rooted in the Schr{\"o}dinger dynamics with an isotropic quadratic dispersion, as commonly realized in semiconductor quantum wells acting as soft-wall billiards~\cite{Nakamura_book}. However, not all quantum devices obey the Schr\"odinger equation; some of them obey the Dirac equation. A distinct class of billiards arises for relativistic wave equations with linear dispersion, first exemplified by the neutrino billiard of Berry and Mondragon~\cite{berry1987neutrino}. On the other hand, low-energy quasiparticles in a monolayer graphene (MLG) cavity emulate the physics of the Dirac equation~\cite{ponomarenko2008chaotic,condado2025exactly,huang2018relativistic}, in which pseudospin, valley degrees of freedom, and relativistic boundary conditions fundamentally modify the correspondence between classical dynamics and quantum properties. As a result of being more accessible than a hypothetical neutrino billiard, an extensive amount of investigations has focused on MLG devices to examine the characteristics of a pseudo-relativistic quantum system, such as Klein tunneling~\cite{Beenakker_rev.mod.phys_80_1337_2008}, the creation and probing of whispering gallery mode resonators~\cite{zhao2015creating}, spectral statistics and wavefunction properties~\cite{geim2007rise,castro2009electronic,PhysRevB.77.085423} and recently the direct visualization of relativistic quantum scars~\cite{ge2024direct}.

Beyond the extensive interest in MLG~\cite{castro2009electronic,DasSarma_rev.mod.phys_83_407_2011,Basov_rev.mod.phys_86_959_2014}, bilayer graphene (BLG) devices have emerged as a highly versatile program in its own right~\cite{mccann2013electronic}. Recent advances have demonstrated exceptional material quality and electrostatic control~\cite{Tong_nano.lett_21_1068_2021,Banszerus_nano.lett_20_7709_2020}, together with a wealth of novel physical phenomena~\cite{Banszerus_nat.commun_13_3637_2022,Garreis_phys.rev.lett_126_147703_2021,Garreis_nat.phys_30_428_2024,Garreis_phys.rev.research_5_013042_2023,Kurzmann_nat.commun_12_6004_2021,Kurzmann_phys.rev.lett_123_026803_2019,Tong_phys.rev.lett_133_017001_2024}. For example, an out-of-plane electric field induces a tunable band gap by breaking layer symmetry, enabling electrostatically defined confinement~\cite{mccann2013electronic, McCann_phys.rev.lett_96_086805_2006, Min_phys.rev.b_75_155115_2007, castro2009electronic, Zhang_nature_459_820_2009}. Alternatively, the necessary confinement can be created through lithographic methods [see, e.g., Refs.~\cite{Bell_nanotechnology_20_455301_2009,Cong_j.phys.chem.c_113_6529_2009, rubio_phys.chem.chem.phys_19_8061_2017}], where the BLG is either cut or grown into a flake of desired geometry, producing hard-wall confinement reminiscent of a quantum billiard. In contrast to MLG, BLG lacks relativistic Dirac dynamics at low energies but instead hosts distinct phenomena, including Zitterbewegung-like motion arising from interlayer coupling~\cite{PhysRevEconductance_2015}. Even though the low-energy dispersion is quadratic, it is strongly modified by trigonal warping~\cite{mccann2013electronic}.

The pronounced role of trigonal warping in BLG has been demonstrated under a wide range of conditions~\cite{Garreis_phys.rev.lett_126_147703_2021,Gold_phys.rev.lett_127_046801_2021,Kechedzhi_phys.rev.lett_98_17680_2007,Kaladzhyan_phys.rev.b_104_235425_2021,Garreis_nat.phys_30_428_2024}, with scanning tunneling microscopy providing quasi-direct access to its anisotropic Fermi surface. This anisotropy renders the electronic dynamics highly sensitive to electrostatic gating~\cite{ohta2006controlling,oostinga2008gate} and strain~\cite{PhysRevB.84.041404}, as well as to lattice orientation. Notably, gate tuning alone can induce a coexistence of regular and chaotic dynamics even in highly symmetric BLG quantum dots~\cite{seemann2023gate}. Compared to MLG and common two-dimensional electron gas structures, trigonal warping allows the semiclassical dynamics to be tuned with far greater flexibility. In fact, semiclassical ray-dynamics and Poincaré analyses have revealed phase-space structures and periodic-orbit stability properties that are qualitatively distinct from those of conventional billiards~\cite{seemann2023gate,seemann2025complex,rv22-w3p8}. Furthermore, the same anisotropy underlies electron-optical effects such as Veselago lensing, underscoring the unusual transport properties of BLG cavities~\cite{cheianov2007focusing}.

While the features of BLG have motivated investigation in transport and device physics~\cite{PhysRevB.81.195406,san2014stacking}, the implications remain largely uncharted in the context of quantum chaos, furthermore almost entirely confined to  semiclassical perspective. The quantum mechanical properties of BLG cavities, such as their eigenstate morphology, spectral statistics, and the extent to which the semiclassical features persist or break down at the quantum level, are essentially unknown. Addressing these issues is crucial for understanding BLG cavities not only as electron-optical devices~\cite{cheianov2007focusing} but also as a platform for systematically investigating quantum-chaotic phenomena in strongly anisotropic materials.

Here, we demonstrate that boundary–lattice misalignment within BLG cavities induces a transition from the near-integrability to chaos, reflected in both eigenvalue statistics and eigenstate anatomy, while semiclassical ray dynamics exposes the classical phase-space restructuring responsible for this behavior that aligns with the prior work by Hentschel and co-workers~\cite{hentschel2025ray,seemann2025complex}. In particular, we investigate the interplay of the discrete nature of the lattice and the cavity geometry, going beyond widely used continuum models, such as utilized in Ref.~\cite{rv22-w3p8}. Taken all together, we present a step toward the systematic investigation and design of chaos in BLG–based devices.   

\begin{figure}[t]
  \centering
  \includegraphics[width=0.95\columnwidth]{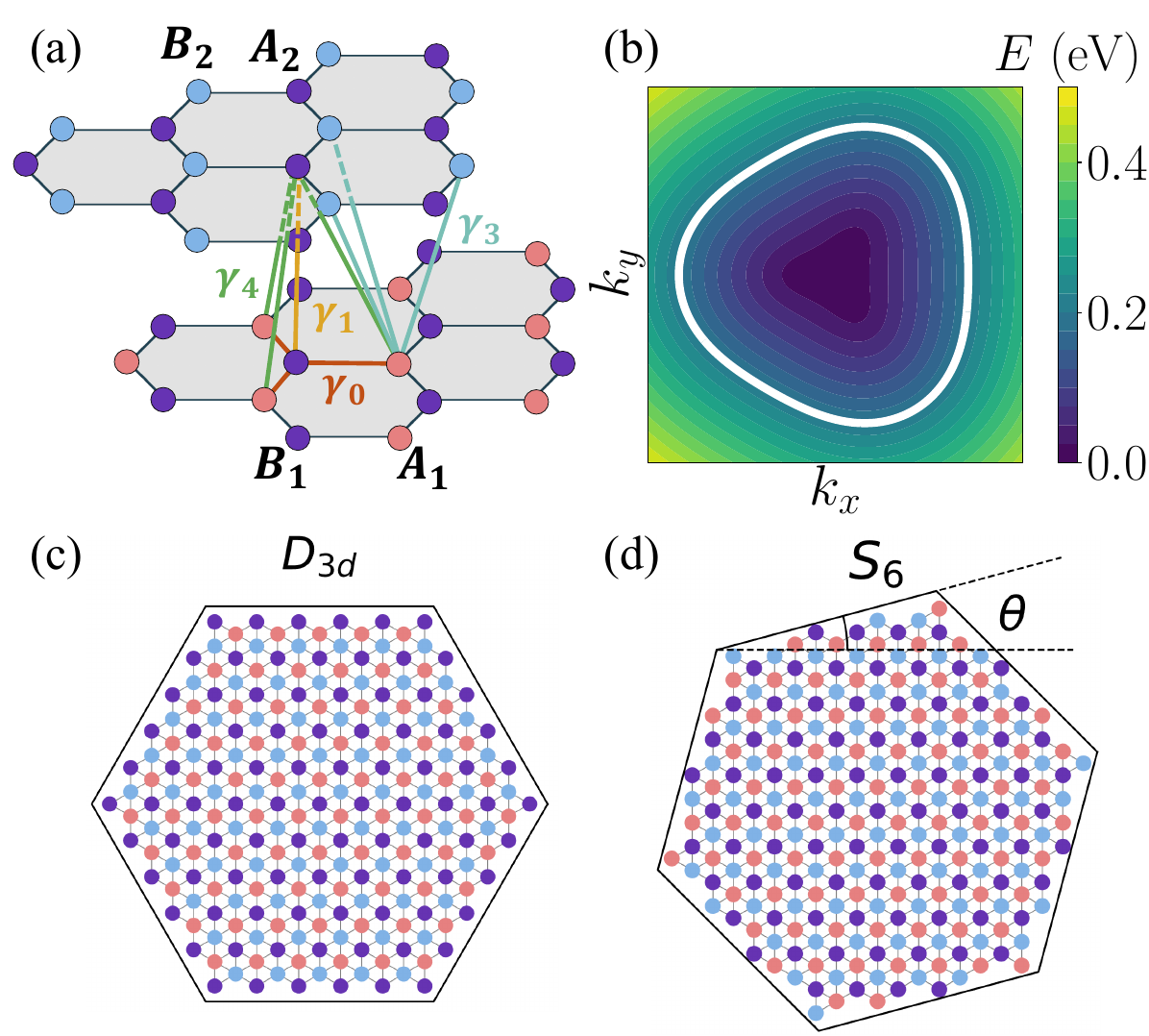}

  \caption{\raggedright
Illustration of the tight-binding model, Fermi surface, and cavity geometry.
    (a) Schematic diagram of the BLG tight-binding model showing atoms and hopping. 
    (b) K valley of BLG's band structure, where the white curve represents the Fermi surface at $E=0.2\,\mathrm{eV}$, showing trigonal warping due to interlayer hopping $\gamma_3$. 
    (c) and (d) show diagrams of BLG cut into a hexagonal shape, where purple indicates the dimer atoms $B_1$ and $A_2$, i.e., the vertically aligned sites that are directly coupled between the two layers, and red and blue indicate  the non-dimer atoms $A_1$ and $B_2$.
    In (c), the boundary is aligned with the inner lattice, and the point group of the atoms is $D_{3d}$.  
    In (d), the boundary is rotated by $15^\circ$ relative to the inner lattice, and the point group of the atoms is $S_6$. The cavities we study contain millions of atoms. 
  }
  \label{FIG1}
\end{figure}

\section*{Model} 

We investigate an AB-stacked BLG cavity with regular hexagonal geometry, in order to elucidate the influence of trigonal warping and lattice–boundary misalignment. More specifically, the electronic structure of this cavity is described by the standard Slonczewski–Weiss–McClure tight-binding Hamiltonian~\cite{mccann2013electronic}:
\begin{align}
  \mathcal{H} &= 
    \sum_{\langle i,j\rangle_{\mathrm{intra}}} 
        \gamma_{0}\, c_{i}^{\dagger} c_{j}
    + \sum_{\langle B_{1}, A_{2}\rangle}
        \gamma_{1}\, c_{B_{1}}^{\dagger} c_{A_{2}}^{}+ \sum_{\langle A_{1}, B_{2}\rangle}
        \gamma_{3}\, c_{A_{1}}^{\dagger} c_{B_{2}}^{}\notag\\
    &
    + \sum_{\langle A_{1}, A_{2}\rangle}
        \gamma_{4}\, c_{A_{1}}^{\dagger} c_{A_{2}}^{}
    + \sum_{\langle B_{1}, B_{2}\rangle}
        \gamma_{4}\, c_{B_{1}}^{\dagger} c_{B_{2}}^{}
    + \text{H.c.},
    \label{eq:TB}
\end{align}
which includes the dominant intralayer hopping $\gamma_0$, the interlayer dimer coupling $\gamma_1$, and the skew interlayer hoppings $\gamma_3$ and skew interlayer hopping between same-sublattice sites $\gamma_4$. In the Hamiltonian above, $c_i^\dagger$ ($c_i$) creates (annihilates) an electron on site $i$, and $\langle\cdot,\cdot\rangle$ denotes the corresponding in-plane or interlayer nearest-neighbor pairs as indicated in Fig.~\ref{FIG1}(a). Furthermore, $A_1$ and $B_1$ in Fig.~\ref{FIG1}(a) refer to the two sublattice atoms in the lower graphene layer, while $A_2$ and $B_2$ refer to the two sublattice atoms in the upper layer. Non-dimer sites in the top and bottom layers are shown in blue and red, respectively, while vertically aligned interlayer dimer sites are marked in purple.

The hopping parameters are chosen in accordance with established literature~\cite{mccann2013electronic} (see \textit{Materials and Methods} for details). Whereas dominant in-plane hopping $\gamma_0$ defines the Dirac-like dispersion within each layer and the vertical interlayer coupling $\gamma_1$ hybridizes the dimer sites ($B_1–A_2$) yielding the characteristic low-energy parabolic bands, we are most interested in the skew parameter $\gamma_3$ which introduces trigonal warping and hence breaks the isotropy of the energy spectrum near the neutrality point. Fig.~\ref{FIG1}(b) shows a trigonally warped Fermi surface of BLG stemming from the non-dimer coupling parametrized by $\gamma_3$. Finally, same-sublattice interlayer hopping $\gamma_4$ couples $A_1$–$A_2$ and $B_1$–$B_2$ sites, lifting up electron–hole symmetry and only slightly modifying the band curvature.

We focus on the energy window $\Delta E = 0.18$–$0.25\,\mathrm{eV}$, where the Fermi surface exhibits pronounced trigonal warping. As displayed in Fig.~\ref{FIG1}(b), the 
$K$-valley Fermi surface becomes strongly anisotropic at energies a few hundred meV above the quadratic band-touching point. Within this energy window, the typical wavelength of a quantum state, $\lambda\approx20\,\mathrm{nm}$, remains much larger than the lattice constant $a\approx 0.246 \, \mathrm{nm}$. Moreover, the characteristic system size $L \approx 800\,\mathrm{nm}\gg \lambda$ places the studied system firmly in the semiclassical regime, justifying the semiclassical analysis employed below. Crucially, we have verified that both the numerical results and the corresponding conclusions below are qualitatively robust against variations in system size (the analysis is carried out in {\it Supplementary Material}).

To directly probe the interplay between the underlying and cavity boundary, we rotate the original hexagonal BLG cavity shown in Fig.~\ref{FIG1}(c) by an angle $\theta$, as demonstrated in Fig.~\ref{FIG1}(d). However, the sixfold rotational symmetry of the lattice renders the atomic configuration $60^\circ$-periodic in the rotation angle $\theta$. Furthermore, since the configurations at the angles $\theta$ and $60^\circ-\theta$ are equivalent, it is hence sufficient to restrict our consideration to the interval $0^\circ \le \theta \le 30^\circ$. The limiting cases $0^\circ$ and $30^\circ$ structurally correspond to boundary-aligned zigzag and armchair terminations, respectively; whereas intermediate angles produce misaligned edges with mixed zigzag–armchair character.

\section*{Results}

We compute eigenvalues and eigenstates of the Hamiltonian in \eqref{eq:TB} for the distinct angles $0^\circ \le \theta \le 30^\circ$, and then quantify the degree of chaos within the energy window $\Delta E$. The simulation details are provided in \textit{Materials and Methods} and \textit{Supplementary Material}. However, before delving into these results, we briefly highlight two underlying key aspects about the chosen system. 

First, we address the symmetry of the system (see \textit{Materials and Methods} for details). In the case of the unrotated cavity ($\theta = 0$) shown in Fig.~\ref{FIG1}(c), the boundary is aligned with the crystalline axes, and the Hamiltonian respects the $D_{3d}$ point-group symmetry, which further decomposes into the six irreducible representations $A_{1g}$, $A_{2g}$, $A_{1u}$, $A_{2u}$, $E_g$, and $E_u$. By contrast, for a representative rotated configuration $(\theta = 15^\circ)$ presented in Fig.~\ref{FIG1}(d)), the boundary orientation breaks the original mirror symmetries, downgrading the point group to $S_6$, with four sectors $A_g$, $A_u$, $E_g$, and $E_u$.

Second, we note that a classical hexagonal billiard is an archetypal example of a pseudointegrable system~\cite{richens1981pseudointegrable,lozej2024intermediate}: its interior angles are rational multiples of $\pi$, so trajectories evolve on higher-genus phase-space surfaces rather than on invariant tori as in a fully integrable system. On the quantum side, such pseudointegrable billiards display hybrid spectral statistics, interpolating between regular and chaotic behavior, and are frequently characterized by semi-Poisson distribution~\cite{berry1984semiclassical,PhysRevLett.62.2769,bogomolny1999models,giraud2004intermediate, biswas1990quantum} (see also \textit{Materials and Methods}). We note, however, that this behavior is not universal and can depend on the specific geometry. More broadly, different polygonal billiards may display distinct spectral statistics. For instance, GOE-type statistics has been reported for certain triangular billiards with all irrational angles~\cite{lozej2022quantum}. Moreover, our BLG cavity differs from these idealized classical and quantum billiards due to trigonal warping and the discrete lattice, which regularizes the sharp boundary corners that underlie classical pseudointegrable dynamics. Therefore, it is not \emph{a priori} clear whether pseudointegrability carries over to our system. As we show next, the BLG cavity inherits this feature to some degree, and we also explore its manifestations in the eigenfunction profiles, a much less studied aspect compared to spectral statistics.
\subsection*{Level statistics} 
\begin{figure}[t]
  \centering

  \begin{tikzpicture}
    \node[inner sep=0] (img) {\includegraphics[width=0.99\columnwidth]{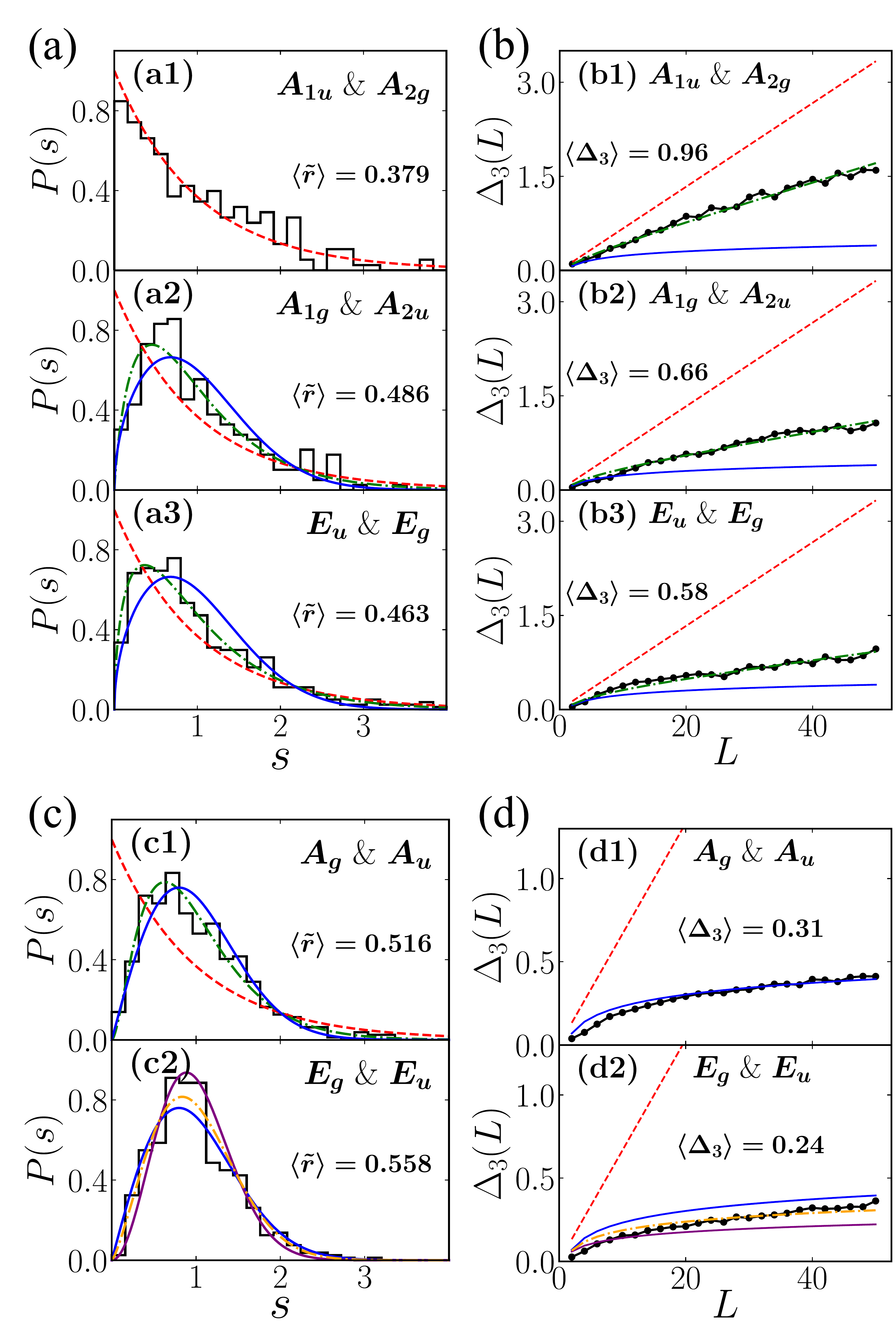}};
  \end{tikzpicture}

  \caption{\raggedright
Level Statistics of unrotated cavities and rotated BLG cavities. (a) and (b) show the level spacing statistics and spectral rigidity for the unrotated cavity, with the corresponding $\langle \tilde{r} \rangle$ values and averaged rigidity indicated in each panel. The red dashed line denotes the Poisson distribution, the blue solid line represents the Wigner–Dyson distribution, and the green dash-dotted line shows the fitted semi-Poisson-like distribution.
(c) and (d) present the same analyses for the cavity rotated by $15^\circ$. The purple solid line represents the GUE distribution, and the orange dash-dotted line denotes the fitted GOE–GUE distribution.}
  \label{FIG2}
\end{figure}

We begin by estimating the degree of chaos in terms of the nearest-neighbor level spacing distribution and the spectral rigidity $\Delta_3(L)$ which are computed within each symmetry-resolved sector, probing the short- and long-range spectral correlations, respectively. Additional quantification is provided by the average gap ratio $\langle \tilde r \rangle$ and the average spectral rigidity $\langle \Delta_3 \rangle$. Figure.~\ref{FIG2} summarizes these measures for the unrotated cavity and for a cavity rotated by $\theta = 15^\circ$. We further classify the (sub)spectra following the spirit of the BGS conjecture~\cite{bohigas1984characterization}: a regular spectrum is associated with Poisson statistics, whereas a chaotic spectrum conform to the Gaussian Orthogonal Ensemble (GOE) or Gaussian Unitary Ensemble (GUE), depending on the presence or absence of time-reversal symmetry, respectively. The latter is also referred as Wigner-Dyson behavior. Our quantum-chaos estimations are detailed in \textit{Materials and Methods}.

The unrotated hexagonal cavity exhibits a pronounced sector-to-sector separation: As seen in Figs.~\ref{FIG2}(a) and (b), two one-dimensional sectors, $A_{1u}$ and $A_{2g}$, are nearly uncorrelated (Poisson-like), whereas the remaining sectors display intermediate statistics close to semi-Poisson, consistent with behavior reported for some pseudointegrable polygonal billiards \cite{PhysRevE.47.54}. This sector dependence reflects the different boundary constraints imposed by the symmetry projections, which can lead to distinct effective dynamics within the same device.

The Poisson-like behavior of the $A_{1u}$ and $A_{2g}$ sectors originates from a symmetry reduction of the boundary-value problem.
In this sector, the parity constraints enforce nodal lines that act as effective Dirichlet boundaries, so the hexagon reduces to an equilateral-triangle fundamental domain (see Fig.~S2 in \textit{Supplementary Material}).
Since the equilateral triangle is integrable, this sector naturally exhibits near-Poisson statistics, as expected based on the standard quantum billiard correspondence. Atomistic zigzag details do not change this large-scale geometry.

Figure~\ref{FIG2}(a) shows the symmetry-resolved statistics for the unrotated cavity within the window $0.18$--$0.25\,\mathrm{eV}$ entailing roughly 3000 eigenstates. The $A_{1u}$ and $A_{2g}$ sectors yield the same value $\langle\tilde r\rangle=0.379$, close to the Poisson limit of $\langle\tilde{r} \rangle_{\textrm{P}} \approx 0.386$, which indicates near-integrability of these sectors. In contrast, the $A_{1g}$ and $A_{2u}$ sectors show stronger level repulsion with $\langle\tilde r\rangle=0.486$, and the $E_u$ and $E_g$ sectors exhibit intermediate statistics with $\langle\tilde r\rangle=0.463$.
Both of these cases are well captured by semi-Poisson-like fits with $\beta=0.83$ and $0.57$, respectively, which are introduced in \textit{Materials and Methods}. In addition, the corresponding rigidity curves in Fig.~\ref{FIG2}(b) follow the same hierarchy, but they appear slightly more Wigner–Dyson–like than suggested by the gap ratio $\langle\tilde r\rangle$. This kind of mismatch is common in the semi-Poisson–like intermediate regime where short-range and long-range correlations can behave differently~\cite{bogomolny2001short}, in line with results reported in a generic framework of mixed dynamics~\cite{persson_phys.rev.e_52_148_1995}. We further observe that the $E$ sectors are systematically more GUE-like than the $A$ sectors, while the latter remain closer to GOE behavior. In other words, time-reversal invariance is lost within the E subspaces, whereas the A subspaces remain time-reversal symmetric~\cite{PhysRevE.104.064211}.

The rotated cavity behaves qualitatively differently from its unrotated counterpart in respected to its unrotated counterpart. As illustrated by Figs.~\ref{FIG2}(c) and (d) for $\theta=15^\circ$,
the $A$ sector remains intermediate with $\langle\tilde r\rangle=0.516$, while the $E$ sector shifts toward Wigner--Dyson correlations with $\langle\tilde r\rangle=0.558$ and a strongly suppressed exponential tail. This bearing is well characterized by a mixed GOE--GUE interpolation~\cite{mehta2004random} described in \textit{Materials and Methods}. It demonstrates that lattice--boundary misalignment lifts up the near-integrability present at $\theta=0^\circ$, and generally drives the spectrum toward fully chaotic statistics. Consistently, the spectral rigidity mirrors the gap-ratio trend: the average rigidity $\langle\Delta_3\rangle$ drops toward the GOE/GUE limits under the considered rotation, attaining values of $0.31$ in the sector $A$ ($\langle \Delta_3 \rangle_{\rm GOE} \approx 0.319$) and $0.24$ in the sector $E$ ($\langle \Delta_3 \rangle_{\rm GUE} \approx 0.184$), which reaffirms fact on the subspace time-reversal symmetry above.

\begin{figure}[t]
  \centering
  \begin{tikzpicture}
    \node[inner sep=0] (img) {\includegraphics[width=0.95\columnwidth]{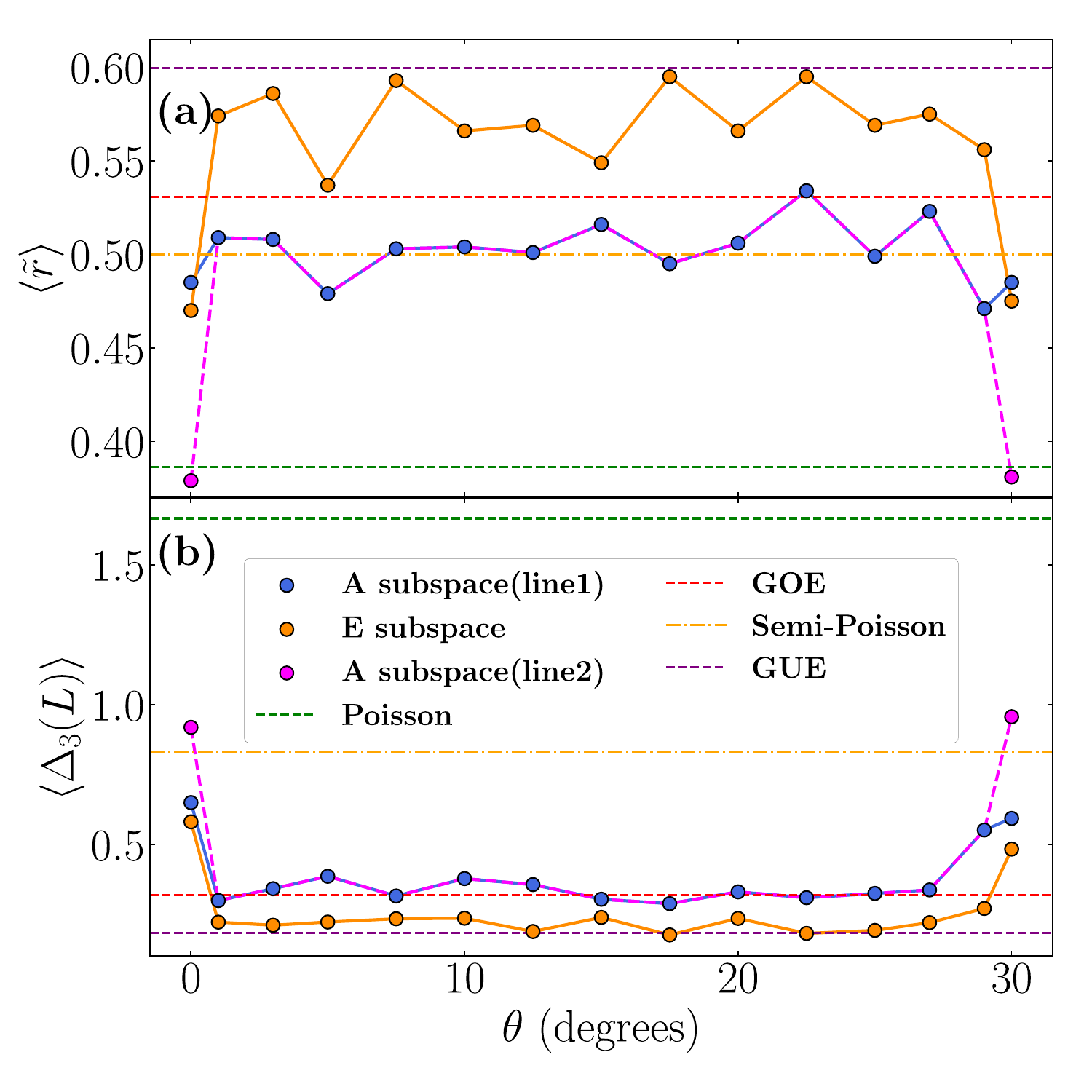}};
  \end{tikzpicture}

  \caption{\raggedright
Angle dependence of $\langle \tilde r \rangle$ and $\langle \Delta_3 \rangle$ on $\theta$. (a) and (b) present the dependence of the \( r \)-value and the average spectral rigidity on the rotation angle~\( \theta \) for the \( A \) and \( E \) subspaces, respectively. In the \( A \) subspaces, two lines are shown: for the unrotated cases \( \theta = 0^\circ \) and \(30^\circ\), the blue solid line corresponds to the pseudointegrable sector, while the purple dashed line represents the integrable sector. Horizontal dashed lines indicate reference values for Poisson, semi-Poisson ($\beta=1$), GOE, and GUE statistics.}
  \label{fig:FIG3}
\end{figure}

Figure~\ref{fig:FIG3} summarizes the evolution across the rotation angle $\theta$. In the $A$ sector, two branches are observed, as shown in Fig.~\ref{fig:FIG3}(a), arising from a symmetry change induced by rotation. For intermediate angles $0^\circ < \theta < 30^\circ$, the two branches (lines 1 and 2) overlap, indicating semi-Poisson–like behavior. At the commensurate alignment of the BLG lattice occurring when $\theta=0^\circ$ and $\theta = 30^\circ$, a clear distinction nevertheless emerges: while the branch corresponding to line 1 remains semi-Poisson–like, line 2 becomes Poissonian.
In the $E$ sector, the gap ratio $\langle\tilde r\rangle$ ncreases markedly from the semi-Poissonian values at $\theta=0^\circ$ and $\theta = 30^\circ$ to the GOE/GUE range over intermediate angles. 

As shown in Fig.~\ref{fig:FIG3}(b), the averaged rigidity $\langle\Delta_3\rangle$, which probes long-range spectral correlations, exhibits the similar overall trend as the gap ratio exploring the short-range spectral correlations. However, in contrast to the gap-ratio analysis, both branches of the $A$ sector fall on the GOE line rather than the semi-Poissonian one, while the $E$ sector clearly aligns with the GUE limit instead of interpolating between GOE and GUE. This provides a clearer manifestation of the differing time-reversal symmetry properties discussed earlier for these subspaces. Taken together, these results of gap ratio and spectral rigidity indicate that the chaos induced by the cavity rotation is robust and does not require fine-tuning of the rotation angle $\theta$.

\begin{figure*}[t]
  \centering
  \begin{tikzpicture}
    \node[inner sep=0] (img) {\includegraphics[width=0.99\textwidth]{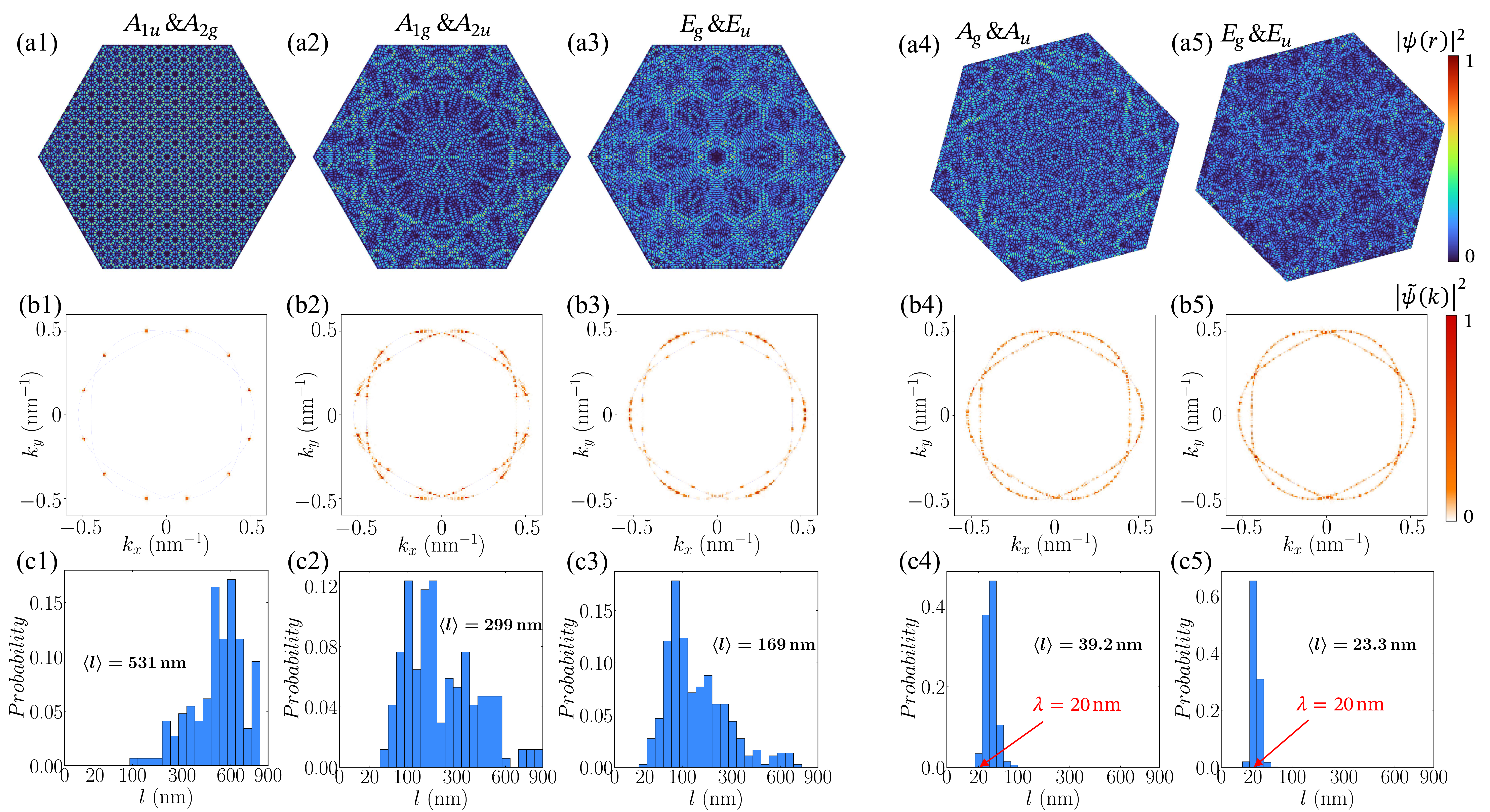}};
  \end{tikzpicture}

  \caption{\raggedright
The wavefunctions and their associated statistical properties. (a1)–(a3) representative probability density distributions $|\psi(\mathbf{r})|^2$ of eigenstates in unrotated cavities for different symmetry subspaces. (a4) and (a5) the corresponding results for the rotated case. (b1)–(b5) the associated distributions $|\tilde{\psi}(\mathbf{k})|^2$ in momentum space with trigonal warped Fermi surface at $K$ and $K'$. (c1)–(c5) the statistical behavior of the correlation length $l$ for the unrotated and rotated cavities across the symmetry subspaces.
  }
  \label{FIG4}
\end{figure*}

\subsection*{Eigenstates}

To further elucidate the effect of lattice–boundary misalignment manifested as the spectral crossover discussed above, we also investigate the anatomy of wavefunctions as the cavity is rotated. As a general assisting principle, we assume that the structure of eigenstates reflects the nature of the underlying classical dynamics: a state of integrable systems displays regular standing-wave patterns, whereas generic states in chaotic systems follow the Berry conjecture~\cite{berry1977regular}, but may also contain non-ergodic patterns, such as quantum scars associated with unstable classical orbits~\cite{mehta2004random,berry1977regular,heller1984bound}. Pseudointegrable systems, by contrast, are expected to interpolate between these limits and often possess symmetry-dependent eigenstate structures~\cite{biswas1990quantum}.

In particular, we quantify changes in the eigenstate profiles under rotation employing an autocorrelation function, from which we extract a correlation length $l$ as a measure of spatial coherence (see \textit{Materials and Methods}). Basically, following our guiding philosophy, smaller values of $l$ indicate more chaotic, weakly correlated states, whereas a larger $l$ is associated with more regular behavior. For comparison, the random-wave model inspired by the Berry conjecture predicts the coherence length $l$ to be equal to the wavefunction de Broglie wavelength $\lambda$, which in our system is $\lambda \sim 20\, \textrm{nm}$ at energies around $0.2\,\mathrm{eV}$. To quantify the degree of chaos as a function of rotation angle, we compute statistical averages of coherent lengths over approximately 1000 eigenstates in this energy neighborhood.

Figure~\ref{FIG4} summarizes our findings on this eigenstate analysis for both the unrotated cavity and the cavity rotated by $\theta = 15^{\circ}$. We illustrate representative eigenstates from each symmetry subspace of the considered system. The coordintate-space probability densities are shown in Figs.~\ref{FIG4}(a1–a5), while Figs.~\ref{FIG4}(b1–b5) display the corresponding momentum-space distributions that are obtained via the Fourier transform $\tilde{\psi}(\mathbf{k}) = \int d^{2}\mathbf{r}\, e^{-i\mathbf{k}\cdot\mathbf{r}}\, \psi(\mathbf{r})$. Furthermore, the corresponding trigonally-warped Fermi surfaces in the $K$ and $K´$ valleys are indicated by the two blue curves. The statistical analysis of the correlation length is outlined in Figs.~\ref{FIG4}(c1–c5).

For the unrotated cavity, the $A_{1u}$ and $A_{2g}$ sectors display regular standing-wave patterns
[Fig.~\ref{FIG4}(a1)], characteristic of integrable eigenstates. Correspondingly, their momentum-space densities are localized at discrete points on the Fermi surface [Fig.~\ref{FIG4}(b1)], which evinces the dominance of only a few plane-wave components in the eigenfunction. On the other hand, the eigenstates in the $A_{1g}$ and $A_{2u}$, $E_u$ and $E_g$ sectors manifest more irregular coordinate-space features  [Figs.~\ref{FIG4}(a2,a3)], accompanied by broader momentum-space distributions involving a larger set of relevant wavevectors [Figs.~\ref{FIG4}(b2,b3)], a hallmark of their pseudointegrability.

In contrast to the unrotated case, the representative eigenstates of the rotated cavity [Figs.~\ref{FIG4}(a4,a5)] display highly irregular density profiles in the coordinate space. Moreover, we observed no clear signatures of periodic orbits or other type of scars, such as superscars~\cite{2004PRLBogomolny,2006PRL_Superscar_Bogomolny} or variational~\cite{keski-rahkonen_phys.rev.b_97_094204_2017, keski-rahkonen_j.phys.conden.matter_31_105301_2019, keski-rahkonen_phys.rev.lett_123_214101_2019}. The corresponding momentum-space densities [Figs.~\ref{FIG4}(b4,b5)] form nearly continuous distributions along the trigonal-warped Fermi surface, which indicates random-wave–like behavior. However, this differs from Berry’s random-wave model, where constitute wave vectors are isotropically distributed, as the random waves here are constrained to the highly warped Fermi surface.

The eigenstate transition to chaos is well captured by our statistical analysis of the correlation length. In the unrotated cavity, all symmetry subspaces yield average correlation lengths exceeding the corresponding de Broglie wavelength $\lambda \sim20\,\mathrm{nm}$ [see Figs.~\ref{FIG4}(a1--a3)], hence indicating the lack of chaos. The integrable $A_{1u}$ and $A_{2g}$ sectors hold the largest value, $\langle l\rangle = 531\,\mathrm{nm}$, whereas the pseudointegrable $A_{1g}$ and $A_{2u}$, $E_u$ and $E_g$ sectors show smaller averages of $\langle l\rangle = 299\,\mathrm{nm}$ and $169\,\mathrm{nm}$. Meanwhile, for a cavity rotated by $\theta = 15^\circ$, the average correlation length in both the $A$ and $E$ sectors dramatically relative to the unrotated case [see Figs.~\ref{FIG4}(c4--c5)], reaching $\langle l \rangle = 39.2\,\mathrm{nm}$ and $\langle l \rangle = 23.3\,\mathrm{nm}$, respectively. Notably, in the $E$ subspace, the correlation lengths of most eigenstates cluster around the de Broglie wavelength, signaling the emergence of spatially uncorrelated, random-wave–like states. In the \textit{Supplementary Material}, we provide an extended statistical analysis of the correlation length as a function of the rotation angle $\theta$. These results demonstrate that rotating the BLG cavity induces chaos by markedly suppressing spatial correlations and increasing the number of contributing plane-wave components in the eigenstates, paralleling to the emergence of chaos in the level statistics. 

\subsection*{Semiclassical dynamics}

While the eigenlevel statistics reveal the emergence of chaos attributed to the lattice–boundary misalignment under cavity rotation, the eigenstate analysis further identifies trigonal warping as an additional contributing factor. To gain intuitive insight into its effect on the BLG cavity dynamics, we complement the tight-binding analysis with a simplified semiclassical ray model. 

To this end, we adopt a continuum bulk description based on the standard four-band Hamiltonian for AB-stacked bilayer graphene~\cite{mccann2013electronic}, which serves here as a minimal model of warping-induced anisotropy in the bulk dispersion. Within this framework, the geometric effects of trigonal warping are captured through an anisotropic group velocity. The model deliberately neglects atomistic edge details, and therefore provides a heuristic description of the dynamical consequences arising from the misalignment between the cavity boundary and the anisotropic Fermi surface. In particular, the cavity can be viewed as an anisotropic Minkowski billiard~\cite{gutkin2002billiards}: particle ray-trajectories propagate with the band velocity $\mathbf{v}(\mathbf{k})=\nabla_{\mathbf{k}}E(\mathbf{k})$ along a warped constant-energy contour. Boundary reflections are implemented by enforcing conservation of energy and of the tangential momentum component (see \textit{Supplementary Material} for details). Unlike conventional specular reflection, this rule is momentum dependent and strongly anisotropic. Nevertheless, because the warped Fermi contour remains strictly convex within the considered energy window, the resulting reflection map is smooth and area-preserving.

\begin{figure}[t!]
  \centering
  \begin{tikzpicture}
    \node[inner sep=0] (img) {\includegraphics[width=0.9\columnwidth]{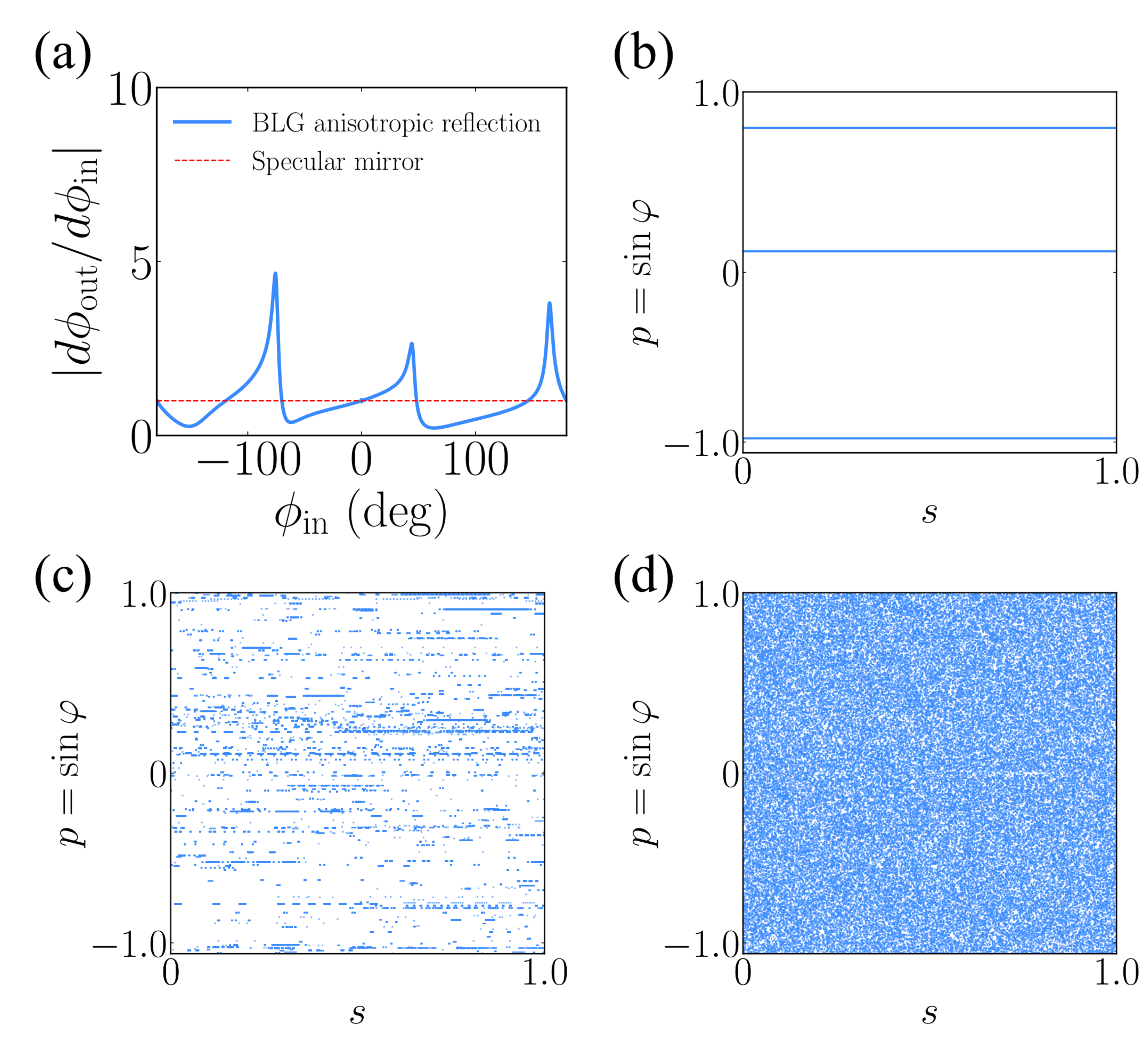}};
  \end{tikzpicture}
  \caption{\raggedright
(a) The angular magnification $|d\phi_{\rm out}/d\phi_{\rm in}|$ for BLG’s anisotropic reflection (blue) compared with the constant unit magnification of specular mirror reflection (red dashed).  The variations in BLG magnification quantify the underlying warping responsible for the structures in (b) and (c). (b–d) Poincaré sections constructed by recording, for a single trajectory crossing a chosen reference edge, the boundary coordinate $s$ (position along the edge) and the incidence angle $\theta$ (relative to the boundary normal).
(b) With the boundary aligned to crystalline $C_3$ axis, the map collapses onto a few invariant curves characteristic of pseudo-integrable motion.
(c) After rotating the cavity by $15^\circ$, the map spreads quasi-ergodically over the accessible phase space.(d) For comparison, a mirror reflecting irrational hexagon shows uniformly filled Poincaré points, demonstrating true ergodicity without crystalline commensurability. }
  \label{FIG5}
\end{figure}

In the ray dynamics of our BLG cavity, the trigonal warping leads to a momentum-dependent and strongly anisotropic reflection rule. To get measure of this aspect, we assess the angular magnification $M(\phi_{\rm in}) = \left|d\phi_{\rm out}/d\phi_{\rm in}\right|$ shown in Fig.~\ref{FIG5}(a), where $\phi_{\rm in}$ and $\phi_{\rm out}$ denote the incidence and reflection angles measured with respect to the local boundary normal, respectively. In contrast to ideal specular reflection ($M = 1$), the angular magnification $M(\phi_{\rm in})$ exhibits a pronounced dependence on the angle $\phi_{\rm in}$ due to trigonal warping. Albeit the angular magnification displays strong variations and sharp peaks at certain incidence angles, our numerical analysis at the same time yields a vanishing maximal Lyapunov exponent $\lambda_{\max} \simeq 0$; thus, the trigonal warping does not induce exponential sensitivity in the semiclassical dynamics. More generally, because the boundary segments are flat and therefore cannot generate exponential separation of nearby trajectories, the Lyapunov exponent is known to remain zero for polygonal billiard~\cite{gutkin1986billiards}. Nevertheless, the condition $\lambda_{\max} = 0$ does not automatically imply unchanged dynamics. In fact, as we demonstrate below, the anisotropic reflection map can indeed strongly modify the ergodic properties of the system.

We visualize the semiclassical ray dynamics of the cavity through the Poincaré sections shown in Fig.~\ref{FIG5}. These sections are constructed
by recording the arclength coordinate s along a fixed reference edge and the momentum variable
$p = \sin \varphi$ where $\varphi$, where $\varphi$ is the incidence angle, each time a ray intersects that edge over $5\times10^5$ collisions. When the cavity boundary is aligned with a crystalline $C_3$ axis, which corresponds to the unrotated case $\theta = 0$, the warped reflection rule becomes commensurate with the boundary symmetry and effectively reduces to ordinary specular reflection~\cite{seemann2023gate}. As a consequence, successive reflections recur along a small set of preferred directions, repeatedly sampling only a discrete set of incidence angles. The corresponding Poincaré section therefore collapses into a few distinct lines rather than filling the accessible phase space, as seen in Fig.~\ref{FIG5}(b). The motion is confined to these invariant lines, a signature of pseudointegrable dynamics.

Upon rotating the BLG cavity by $\theta = 15^\circ$, the reflection rule associated with the warped Fermi surface becomes incommensurate with the cavity geometry, and the incidence angles no longer recur coherently. The resulting Poincaré section, shown in Fig.~\ref{FIG5}(c), exhibits a dense yet structured phase-space pattern consistent with the quasi-ergodic regime described in Ref.~\cite{wang2022statistical}: trajectories become dense in phase space, but the distribution of points remains markedly non-uniform even after long evolution times. Specifically, the rotated BLG cavity displays bands of enhanced density and filamentary streaks across the section, with pronounced variations in phase-space occupancy. This type of slow and non-uniform filling of phase space provides a classical counterpart to the spectral and eigenstate crossover from pseudointegrability to chaos discussed above. By contrast, a genuinely ergodic billiard would yield Poincaré points that fill the allowed domain nearly uniformly within short times, as illustrated for a mirror-reflecting irrational hexagon in Fig.~\ref{FIG5}(d) (see \textit{Supplementary Material} for further details).


\section*{Discussion and Conclusion}

\subsection*{Semiclassical explanation of quantum chaos}

The engender of chaos upon misalignment between the cavity boundary and the BLG lattice can, to some extent, be understood from a semiclassical viewpoint. We stress, however, that the moniker \emph{chaos} here primarily refers to quantum idiosyncrasies of chaos, as evidenced by Wigner–Dyson level statistics and random-wave–like eigenstates. While we make use of a simplified semiclassical ray-dynamics construction, its purpose is only to elucidate the geometric consequences of the warped dispersion rather than to provide a complete microscopic account. In particular, we underscore the fact that no simple semiclassical framework can fully capture the atomistic boundary physics inherent in the underlying lattice model.

Even so, the semiclassical picture offers an appealing interpretation of the observed transition. For instance, such semiclassical ergodicity, as exhibited by 
the rotated cavity, is known to suffice for generating Wigner–Dyson statistics, even in the absence of classical chaos attributed to positive Lyapunov exponents. Indeed, studies of non-hyperbolic systems, including the quantum triangle map~\cite{wang2022statistical}, triangular cavities~\cite{lozej2022quantum}, and parabolic triangle maps~\cite{degli2005semi}, have demonstrated that classical quasi-ergodicity alone can produce RMT level statistics in the semiclassical limit. These findings accentuate the central role of classical mixing and ergodic properties in shaping quantum spectral behavior.

More generally speaking, in pseudointegrable or quasi-ergodic systems, spectral statistics and eigenfunction measures extracted from a finite energy window can depend sensitively on the rate at which the corresponding classical dynamics mixes within the accessible phase space. The approach to uniform phase-space coverage may be exceedingly slow, so finite-shell quantum diagnostics do not need to coincide with the classical limit received at $t \rightarrow \infty$, as advocated by a shell-by-shell, finite-time correspondence perspective in Ref.~\cite{PhysRevLett.134.130402}. From this viewpoint, the rotated BLG cavity provides a natural example of slow phase-space filling induced by an anisotropic reflection law: trajectories initially populate filamentary bands in the Poincaré section and only gradually disperse over longer times.

We further note that the reflection law in our BLG cavity is itself energy dependent. As the energy window changes, the Fermi-surface geometry evolves and the relevance of the four-band description exploited above may vary, leading to corresponding modifications of the effective semiclassical reflection rule. The observed spectral crossover from pseudo-integrability to RMT statistics and the emergence of random-wave eigenstates should therefore be understood as a qualitative correspondence at the level of a selected energy shell and finite-time classical dynamics, rather than as evidence of strict asymptotic ergodicity. A more systematic shell-by-shell quantum–classical correspondence analysis for anisotropic BLG cavities nonetheless remains an interesting direction for future research.

\subsection*{Edge effect and comparison to MLG}

For a rotated BLG cavity defined on a lattice, we underline the importance to distinguish chaos caused by the geometric incommensurability with the warped Fermi surface from atomistic edge roughness. Rotating the cavity relative to the underlying lattice produces a boundary composed of mixed zigzag and armchair segments, and consequently introduces additional kinks. In MLG, such mixed edges are known to modify intervalley scattering and thereby influence spectral correlations and eigenstate properties in certain cavity geometries~\cite{yu2016gaussian,hagymasi2017interaction}. It is therefore natural to ask whether edge effects could dominate the behavior observed in our BLG cavity.

To clarify this point, we perform two control tests (see \textit{Supplementary Material}). First, we artificially introduce kinks into an unrotated cavity and observe that increasing the number of kinks leads to a gradual rise in the gap ratio $\langle\tilde r\rangle$, implying a progressive enhancement of chaotic spectral features. Second, we rotate an analogous MLG cavity, where trigonal warping is absent at low energies and the symmetry is reduced from $D_6$ to $C_6$ upon rotation. In this case, the gap ratio $\langle\tilde r\rangle$ increases only slightly, and the spectrum remains semi-Poisson–like even up to a rotation angle of $\theta = 10^\circ$. Therefore, we conclude that edge roughness contributes only gradually and in an angle-dependent manner to the spectral correlations, rather than being the primary driver of the observed chaos transition.

In the BLG cavity, by contrast, trigonal warping provides a far more effective mechanism capable of inducing a rapid transition towards chaos. A rotation as small as $\theta = 0.1^\circ$ that corresponds to shifting the lattice by only a single kink and causing negligible geometric deformation is already sufficient to drive the spectrum toward Wigner-Dyson statistics. At larger rotation angles, edge effects become increasingly relevant: the growing armchair content enhances intervalley scattering, and the semiclassical reflection rule ceases to be a controlled approximation~\cite{PhysRevB.84.075468}. We therefore interpret the emergence of chaos in the rotated BLG cavity as the combined result of (i) a dominant trigonal-warping mechanism that promptly destroys commensurability between the boundary and the bulk dynamics, i.e. the mismatch with the warped Fermi surface, and (ii) additional edge-related contributions that gain importance as the rotation angle increases.

An important exception occurs at the special angle $30^\circ$, where the lattice and boundary regain commensurability and edge physics becomes dominant. In this configuration, the spectrum shifts back toward pseudointegrable behavior. The boundary approaches a nearly uniform armchair termination, strengthening intervalley processes. Because our semiclassical model neglects atomistic edge conditions, its bulk-mismatch criterion is not expected to capture this edge-driven symmetry restoration.

As a side note, even though trigonal warping also exists in MLG~\cite{2009relativistic}, it is negligible in the low-energy, Dirac regime relevant here, and the associated anisotropic reflection mechanism is hence strongly suppressed. Most MLG cavities and gate-defined quantum dots operate experimentally near the Dirac point, while the higher-energy regime where warping becomes appreciable is less commonly explored~\cite{ge2024direct,Lee2016}. An exception may arise in the presence of strong spin–orbit coupling, where Rashba interaction can induce trigonal warping already at low energies~\cite{TrigonalWarpingRashbaSOC}. In this regime, however, the Fermi wavelength can approach the lattice scale, rendering lattice effects more pronounced and complicating the semiclassical description.

\subsection*{Experiment relevance}

The BLG cavity model considered in this work is experimentally accessible, as already implied in  the introductory. For example, hexagonal graphene cavities can be realized via CVD growth, and the rotated geometry studied here corresponds to a hexagon with kinks~\cite{2013PNAS_CVD}. Alternatively, such cavities can be fabricated by patterning BLG into discrete shapes to achieve hard-wall confinement~\cite{ponomarenko2008chaotic,barreiro2012quantum,hagymasi2017interaction}, or by employing electrostatic gates to define a soft confinement potential, thus enabling tunable control of cavity occupancy and tunnel barriers~\cite{subramaniam2012wave,goossens2012gate,kurzmann2019charge}. Low-temperature transport measurements, including Coulomb blockade and excited-state spectroscopy, provide access to energy-level statistics and wavefunction characteristics, thereby allowing experimental identification of quantum-chaotic signatures~\cite{banszerus2020single}. Together, these approaches offer complementary routes to investigate and control quantum-chaos features in BLG-based systems.

\subsection*{Future directions}

Future studies of BLG cavities may explore the controlled tuning of quantum behavior through external parameters. A particularly promising direction is the application of an external electric field, which can modify electron dynamics and selectively generate or tailor specific wavefunction patterns. The interplay between the electric field, cavity geometry, and trigonal warping may give rise to a rich landscape of dynamical regimes and quantum phenomena. In particular, it could enable controlled manipulation of superscars in classically pseudointegrable cavities~\cite{2004PRLBogomolny,2006PRL_Superscar_Bogomolny}, and potentially open a route toward realizing variational scarring~\cite{keski-rahkonen_phys.rev.b_97_094204_2017, keski-rahkonen_j.phys.conden.matter_31_105301_2019, keski-rahkonen_phys.rev.lett_123_214101_2019} in this class of systems. Additional tuning knobs, such as an applied magnetic field or an interlayer potential difference, may further control the system by modifying its symmetry properties and internal electronic dynamics

Beyond regular polygons, BLG cavities with curved boundaries, such as the well-known stadium geometry, provide a promising platform for exploring chaos arising from boundary misalignment relative to the trigonally warped Fermi surface. More broadly, our work lays the groundwork for investigating trigonal-warping effects, such as the chaos it can induce, in BLG-based quantum dots subject to smooth confinement. Such systems would not only offer an attractive setting for probing the quantum nature of chaos, but could possible enable direct visualization of Heller-type scars via scanning tunneling microscopy analogous to Ref.~\cite{ge2024direct}, as well as the observation of variational scars in a similar fashion as proposed for the MLG counterpart in Ref.~\cite{keskirahkonen_phys.rev.e_112_L012201_2025}.

\subsection*{Conclusion}

In conclusion, we have demonstrated that misalignment between the BLG lattice and the hexagonal boundary of the cavity fundamentally reshapes both its spectral and eigenstate properties. In particular, rotating the boundary lifts the high-symmetry subspaces present in the unrotated geometry, driving the system towards the RMT limit of chaos. Furthermore, our semiclassical analysis attributes this quantum transition primary to the loss of commensurability between the warped Fermi surface and the polygonal boundary. This mismatch promotes quasi-ergodic phase-space exploration which in turn enables to explain
the emergence of quantum-chaotic spectral and eigenstate characteristics. Taken together, our results show that the interplay between boundary orientation and lattice structure, combined with Fermi-surface anisotropy, offers an effective and experimentally accessible means of engineering and controlling quantum-chaotic features in BLG cavities. These findings bear direct relevance to tunable electron transport in graphene-based and other mesoscopic systems~\cite{Velasco_2012,RevModPhys2015_cao,PhysRevEconductance_2015}.

\section*{Materials and Methods}

\subsection*{Numerical implementation}

All numerical simulations presented in this work are performed utilizing the \texttt{pybinding} Python package, which provides a flexible and efficient framework for tight-binding modeling of two-dimensional materials~\cite{moldovan2020pybinding}. Working with an atomistic tight-binding Hamiltonian on the physical honeycomb bilayer lattice avoids discretizing a continuum model on a Cartesian real-space grid, which can introduce artificial square-lattice ($D_4$) anisotropies that contaminate symmetry-sensitive
spectral statistics \cite{rv22-w3p8}. For the same reason, our simulations do not suffer from spurious fermion-doubling artifacts associated with finite-difference discretizations of continuum Dirac theories \cite{nielsen1981absence}.

\subsection*{Tight-binding model parameters}
For the tight-binding model of BLG, we adopt the commonly used parameters for AB-stacked layers, as reported in Ref.~\cite{PhysRevB.81.195406}. Specifically, the lattice constant for BLG is $a = 0.246~\mathrm{nm}$, and the interlayer distance is $d = 0.335~\mathrm{nm}$. The intralayer nearest-neighbor hopping is $\gamma_0 = 3.16$~eV, the interlayer vertical hopping between dimer sites is $\gamma_1 = 0.381$~eV, and the skew interlayer hopping are $\gamma_3 =0.38$~eV and the same-sublattice interlayer hopping $\gamma_4 = 0.14$~eV. 

The circumradius of the cavity is set to $r = 400\, \textrm{nm}$ that corresponds to a characteristic size $L \approx 800\, \textrm{nm}$. In our numerical calculations, the hexagonal cavities contain on the order of $10^6$ lattice sites, but we are still able to to resolve fully quantum spectrum in this regime where semi-classical or continuum approximations would commonly be applied.

\subsection*{Symmetry analysis}
\label{symmetry}
The symmetry of the tight-binding Hamiltonian differs between the two exemplary configurations with the cavity rotation angles $\theta = 0^{\circ}$ and $\theta = 15^{\circ}$. A proper characterization of their quantum-chaotic nature requires analyzing the eigenvalues and eigenstates independently within each symmetry subspace, as defined below via symmetry analysis.

When the boundary is aligned with the crystalline axes, the tight-binding Hamiltonian respects the full $D_{3d}$ symmetry of the BLG lattice. This group contains a threefold rotation $C_{3}$, inversion $\mathcal{I}$, and three vertical mirror operations $\sigma_v$. In this unrotated case, since $[\mathcal{H},\mathcal{R}_{2\pi/3}]=0$, the eigenstates of $\mathcal{H}$ can be chosen as simultaneous eigenstates of the $120^\circ$ rotation operator $\mathcal{R}_{2\pi/3}$, namely
\begin{equation}
\mathcal{R}_{2\pi/3}\psi_n^{(m)}(\mathbf{r})
= e^{i 2\pi m/3}\psi_n^{(m)}(\mathbf{r}),\quad m=0,\pm1.
\end{equation}
The rotation-invariant sector $m=0$ further splits under inversion
$\mathcal{I}$ and mirror reflection $\sigma_v$ into four one-dimensional
irreducible representations.  
To make these parity properties explicit, we label the corresponding
eigenstates as $\psi_n^{(0,p_i,p_\sigma)}$, where $p_i=\pm 1$ and
$p_\sigma=\pm 1$ denote the eigenvalues under inversion and mirror
operations, respectively. In other words, these states satisfy the following equations,
\begin{equation}
    \mathcal{I}\,\psi_n^{(0,p_i,p_\sigma)}
    = p_i\,\psi_n^{(0,p_i,p_\sigma)}
\end{equation}
and
\begin{equation}
    \sigma_v\,\psi_n^{(0,p_i,p_\sigma)}
    = p_\sigma\,\psi_n^{(0,p_i,p_\sigma)}. 
\end{equation}

In this notation,
$(p_i,p_\sigma)=(+,+)$, $(+,-)$, $(-,+)$, and $(-,-)$ correspond to the
irreducible representations $A_{1g}$, $A_{2g}$, $A_{1u}$, and $A_{2u}$,
respectively. The correspondence between the eigenvalues of the symmetry operations ($m$, $p_i$, and $p_\sigma$) and the irreducible representations is summarized in Table \ref{tab:D3d_S6_standard} (left). 
Instead, when the boundary is rotated by an angle $\theta$, the mirror symmetry is broken, and the symmetry group of the tight-binding Hamiltonian reduces to $S_6$. The corresponding symmetry properties and irreducible representations are listed in Table~\ref{tab:D3d_S6_standard} (right). An extended discussion of the wavefunction symmetry properties can be found in the \textit{Supplementary Material}.

\begin{table}[htbp]
\centering
\scriptsize   
\caption{\raggedright
Symmetry sectors used for $D_{3d}$ (left) 
and $S_{6}$ (right). Listed are the irreducible subspace's dimension,
rotation sector $m$, and inversion/mirror parities $p_i/p_\sigma$.}
\label{tab:D3d_S6_standard}

\setlength{\tabcolsep}{2.5pt}  

\begin{tabular}{@{}c@{}c@{}}

\begin{tabular}{@{}l l c c c c@{}}
\toprule
Group & Irrep & Dim. & $m$ & $p_i$ & $p_\sigma$ \\
\midrule
\multirow{6}{*}{$D_{3d}$} 
 & $A_{1g}$ & $1$ & $0$ & $+1$ & $+1$ \\
 & $A_{2g}$ & $1$ & $0$ & $+1$ & $-1$ \\
 & $A_{1u}$ & $1$ & $0$ & $-1$ & $+1$ \\
 & $A_{2u}$ & $1$ & $0$ & $-1$ & $-1$ \\
 & $E_{g}$  & $2$ & $\pm1$ & $+1$ & $0$ \\
 & $E_{u}$  & $2$ & $\pm1$ & $-1$ & $0$ \\
\bottomrule
\end{tabular}

\hspace{3mm} 

\begin{tabular}{@{}l l c c c c@{}}
\toprule
Group & Irrep & Dim. & $m$ & $p_i$ \\
\midrule
\multirow{4}{*}{$S_6$} 
 & $A_{g}$  & $1$ & $0$ & $+1$\\
 & $A_{u}$  & $1$ & $0$ & $-1$ \\
 & $E_{g}$  & $2$ & $\pm1$ & $+1$ \\
 & $E_{u}$  & $2$ & $\pm1$ & $-1$ \\
\bottomrule
\end{tabular}

\end{tabular}
\end{table}

\subsection*{Level statistics}

To characterize the spectral properties for each orientation angle $\theta$, we compute the ordered eigenvalues $\{E_i\}$ of the tight-binding Hamiltonian and analyze their fluctuations after unfolding.  Nearest-neighbor spacings are defined as $s_i = E_{i+1}-E_i$, and the resulting spacing distribution $P(s)$ is compared with the standard Poisson and Wigner–Dyson forms of random-matrix theory \cite{bohigas1984characterization}. In addition, we evaluate the ratio of consecutive spacings,
$r_i = \frac{s_{i+1}}{s_i}$, $ 
\tilde r_i = \min(r_i,1/r_i)$, which takes the characteristic values $\langle\tilde r\rangle_{\mathrm{P}} \simeq 0.386, \langle\tilde r\rangle_{\mathrm{GOE}} \simeq 0.531,$ and $\langle\tilde r\rangle_{\mathrm{GUE}} \simeq 0.600$ \cite{atas2013distribution}.

Long-range correlations of spectrum are quantified in terms of the spectral rigidity~\cite{dyson1963statistical}, which is defined as
\begin{equation}
\Delta_3(L)=\frac{1}{L}\min_{\{A,B\}}\int_E^{E+L}\!\!\big[N(E')-AE'-B\big]^2\, dE'    
\end{equation}
where $N(E)$ is the unfolded staircase function. Qualitatively speaking, integrable spectra show a linear increase of $\Delta_3(L) \propto L/15$, whereas quantum-chaotic spectra exhibit the logarithmic behavior predicted by RMT. To quantify the degree of chaos in the system, we evaluate the spectral rigidity estimator $\langle \Delta_3(L) \rangle$, obtained by averaging $\Delta_3(L)$ over the range $L \in [10, 40]$. For comparison, the characteristic values of $\langle \Delta_3(L) \rangle$ in the three limiting cases are approximately $\langle \Delta_3 \rangle_{\rm Poisson} \simeq 1.67$, $\langle \Delta_3 \rangle_{\rm GOE} \simeq 0.319$, and $\langle \Delta_3 \rangle_{\rm GUE} \simeq 0.184$.

Since different spectral measures can yield distinct estimates of quantum chaos in systems with mixed regular and chaotic features, as demonstrated in Ref.~\cite{persson_phys.rev.e_52_148_1995}, we employ both nearest-neighbor spacing statistics and spectral rigidity that probe different spectral correlations to properly assess the "chaoticity" of our BLG cavity. Furthermore we also consider a semi-Poisson eigenvalue distribution~\cite{bogomolny2009spectral,bogomolny2001short,bogomolny2004spectral}, which provides an appropriate description of pseudointegrable systems and is defined as
\begin{equation}
P_{\rm SP}(\beta, s) = A_\beta \, s^\beta \, \exp\!\left[-(\beta + 1)s\right],
\label{eq:semipossion}
\end{equation}
where the coefficient $A_\beta$ is given in terms of the gamma function as
\begin{equation}
    A_\beta = \frac{(\beta + 1)^{\beta + 1}}{\Gamma(\beta + 1)}.
\end{equation}
In the limit of RMT, we seek to more clearly differentiate between GOE and GUE statistics. Accordingly, we employ their mixed distribution~\cite{mehta2004random}, given by
\begin{equation}\label{eq:wigner}
P_{\rm WD}(\gamma, s) 
= a_\gamma \, s^\gamma \, \exp\!\left(- b_\gamma s^2 \right), 
\end{equation}
where we have defined 
\begin{equation}
b_\gamma 
= \left[
\frac{\Gamma\!\left(\frac{\gamma+2}{2}\right)}
     {\Gamma\!\left(\frac{\gamma+1}{2}\right)}
\right]^2
\quad \textrm{and} \quad
a_\gamma =
\frac{2\, b_\gamma^{(\gamma+1)/2}}
     {\Gamma\!\left(\frac{\gamma+1}{2}\right)}.
\end{equation}
The limiting cases of GOE and GUE corresponds to the values $\gamma=1$ and $\gamma=2$, respectively. Here, the fitting parameters $\beta$ and $\gamma$ provide additional insight into the quantum-chaotic nature of the system, hence complementing the chaos indicators based on nearest-neighbor statistics and spectral rigidity.

\subsection*{Correlation function analysis}

To quantify the degree of chaos from the viewpoint of the eigenstates, we turn our attention onto the following two-point correlation function:
\begin{equation}
C(\mathbf r,\mathbf r')
= \langle \psi^*(\mathbf r)\psi(\mathbf r') \rangle .
\end{equation}
According to the Berry conjecture, a generic wavefunction $\psi(\mathbf{r})$ of a chaotic system can be approximated as a superposition of plane waves with random phases and a fixed wavenumber $k$, formally $\psi(\mathbf r)=\sum_{n=1}^{N} a_n e^{i\mathbf k_n\cdot \mathbf r}$. In this situation, the correlation function becomes isotropic after ensemble averaging, and takes the universal form of
\begin{equation}
C(r)=J_0(k r),
\end{equation}
where $J_0$ is the zeroth-order Bessel function of the first kind, depending only on the separation
$r = |\mathbf r-\mathbf r'|$.~\cite{berry1977regular}

In contrast, in our model the correlation function is generally anisotropic and depends on both the magnitude and direction of the separation vector.
Therefore, to extract a chaos measure from the anisotropic correlation function,
we first focus on the following correlation function
\begin{equation}
C(r) \equiv 
\big\langle \max_{\hat{\mathbf{r}}'} \left\{ 
C(\boldsymbol{\rho}, \boldsymbol{\rho} + \mathbf{r}') \right\} 
\big\rangle_{\boldsymbol{\rho}}
\end{equation}
where the maximization is performed over all directions
$\hat{\mathbf r}'=\mathbf r'/|\mathbf r'|$ of the displacement vector
$\mathbf r'$, for a fixed separation $|\mathbf r'|=r$, 
and $\langle\cdots\rangle_{\boldsymbol{\rho}}$ denotes an average 
over initial positions $\boldsymbol{\rho}$.
We then utilize this correlation function by fitting it to the form
\begin{equation}
C(r)=J_0\!\left(2\pi\frac{r}{l}\right),
\end{equation}
where $l$ characterizes the spatial correlations of the eigenstates. In a fully chaotic system, the correlation length reduces to the de Broglie wavelength $\lambda$ of the given eigenstate. We stress that the extracted correlation length $l$ serves only as a relative and heuristic measure of eigenstate chaos, understood with respect to the benchmark provided by the Berry conjecture.

\subsection*{Semiclassical ray-dynamics}

Direct time-stepping ray propagation in polygonal billiards with anisotropic dispersion is prone to numerical issues near grazing incidence and near vertices, where corner sensitivity can lead to trajectory-dependent ambiguities. To obtain stable long-time dynamics, we implement the semiclassical motion as a discrete map between successive boundary collisions. We tabulate the relation between the incidence angle $\varphi_{\mathrm{in}}$
and the outgoing angle $\varphi_{\mathrm{out}}$, determined by the conservation of energy and the tangential component $k_t$.
Each table contains about $3\times 10^5$ sampled angles.
This approach encodes the warped reflection law once and for all,
and removes the need to solve for $\varphi_{\mathrm{out}}$ by conducting a local root search at every collision.

Between collisions, the next collision point is found by ray-boundary intersection.
The dynamics is thus an alternation of (i) intersection updates for free flight and (ii) table-based
updates for reflection.
This map-based scheme maintains high angular accuracy even for nearly grazing trajectories and avoids artificial corner sticking and other corner-related numerical artifacts.
As a result, we can compute long trajectories and Poincar\'e sections robustly.

\acknowledgements{This project was supported by the National Science Foundation (Grant No. 2403491). We are thankful for the useful discussions with E. Räsänen, A. M. Bozkur, and J. Velasco Jr. A.M.G. thanks the Studienstiftung des Deutschen Volkes for financial support. }

\section*{Data, Materials, and Software Availability}
All data supporting the findings of this study are provided in the main text and/or the SI Appendix. Tight-binding simulations were performed employing the \texttt{pybinding} package~\cite{moldovan2020pybinding}, with all parameter values specified in the main text and/or the SI Appendix. Semiclassical ray-dynamics simulations were carried out utilizing code deposited in GitHub~\cite{BLG_hexagon_chaos_gitlab}, which also includes the corresponding simulation parameters.

\bibliography{references}

\end{document}


\preprint{Shaping chaos in bilayer graphene cavities}

\title{Supporting Information for \\ Shaping Chaos in Bilayer Graphene Cavities}

\author{Jucheng Lin}
\thanks{These authors contributed equally to this work.}
\affiliation{Department of Physics, Harvard University, Cambridge, Massachusetts 02138, USA}
\affiliation{Department of Chemistry and Chemical Biology, Harvard University, Cambridge, Massachusetts 02138, USA}
\affiliation{Department of Physics and Astronomy, Shanghai Jiao Tong University, Shanghai 200240, China}

\author{Yicheng Zhuang}%

\thanks{These authors contributed equally to this work.}

\affiliation{Department of Physics, Harvard University, Cambridge, Massachusetts 02138, USA}
\affiliation{Department of Chemistry and Chemical Biology, Harvard University, Cambridge, Massachusetts 02138, USA}
\affiliation{Department of Physics, School of Physics, Peking University, Beijing 100871, China}

\author{Anton M. Graf}
\affiliation{Department of Physics, Harvard University, Cambridge, Massachusetts 02138, USA}
\affiliation{Department of Chemistry and Chemical Biology, Harvard University, Cambridge, Massachusetts 02138, USA}
\affiliation{Harvard John A. Paulson School of Engineering and Applied Sciences, Harvard University, Cambridge, Massachusetts 02138, USA}

\author{Eric J. Heller}
\email{eheller@fas.harvard.edu}
\affiliation{Department of Physics, Harvard University, Cambridge, Massachusetts 02138, USA}
\affiliation{Department of Chemistry and Chemical Biology, Harvard University, Cambridge, Massachusetts 02138, USA}

\author{Joonas Keski-Rahkonen}
\affiliation{Department of Physics, Harvard University, Cambridge, Massachusetts 02138, USA}
\affiliation{Department of Chemistry and Chemical Biology, Harvard University, Cambridge, Massachusetts 02138, USA}

\date{\today}

\maketitle
\tableofcontents

\section{Simulation details}

Here, we further describe in detail the methods used in our simulations.
\subsection{Scaling method}

We use the scaling method in the tight-binding model~\cite{PhysRevLett.114.036601}. 
The low-energy continuum Hamiltonian for AB-stacked bilayer graphene in the basis $(A_1, B_1, A_2, B_2)$ is

\begin{equation}
H =
\begin{pmatrix}
\epsilon_{A1} & v \pi^\dagger & -v_4 \pi^\dagger & v_3 \pi \\
v \pi & \epsilon_{B1} & \gamma_1 & -v_4 \pi^\dagger \\
- v_4 \pi & \gamma_1 & \epsilon_{A2} & v \pi^\dagger \\
v_3 \pi^\dagger & -v_4 \pi & v \pi & \epsilon_{B2}
\end{pmatrix},
\label{fourband}
\end{equation}
where the effective velocities are
\[
v = \frac{\sqrt{3} a \gamma_0}{2 \hbar}, \quad
v_3 = \frac{\sqrt{3} a \gamma_3}{2 \hbar}, \quad
v_4 = \frac{\sqrt{3} a \gamma_4}{2 \hbar},
\]
and the momentum operator is
\[
\pi = \xi p_x + i p_y,
\]
with $\xi = \pm 1$ for the $K$ and $K'$ valleys.  

We employ the scaling method to enlarge the distances between atoms, thereby reducing the total number of atoms in our simulation. The scaled lattice constant is given by $a_s = s\,a$, where $s$ is the scaling factor and we take $s=3$ in our work. To preserve the Hamiltonian and the low-energy band structure, the hopping parameters are scaled accordingly:
\[
\gamma_{0s} = \frac{\gamma_0}{s}, \quad
\gamma_{1s} = \gamma_1, \quad
\gamma_{3s} = \frac{\gamma_3}{s}, \quad
\gamma_{4s} = \frac{\gamma_4}{s}.
\]

\subsection{Unfolding method}
To analyze the spectral correlations, we first sort the eigenenergies $\{E_i\}$ in ascending order.
The raw spectrum is then unfolded to remove the global variation of the density of states.
Specifically, we construct the integrated level counting function $N(E)$ and approximate it by a smooth function $\overline{N}(E)$ obtained from a polynomial fit~\cite{Abul_Magd_2014}.
The unfolded energies are defined as
\begin{equation}
    \varepsilon_i = \overline{N}(E_i),
\end{equation}
such that the mean level spacing is unity.
The nearest-neighbor level spacings are given by
\begin{equation}
    s_i = \varepsilon_{i+1} - \varepsilon_i,
\end{equation}
and are subsequently normalized by their average value.
Based on the unfolded spectrum, we compute the nearest-neighbor spacing distribution $P(s)$, the ratio of consecutive spacings $\tilde r = \min(s_i,s_{i-1})/\max(s_i,s_{i-1})$, and the spectral rigidity $\Delta_3(L)$ to characterize the spectral statistics.

\subsection{Sub-sublattice}
\begin{figure}[t]
  \centering
  \includegraphics[width=0.95\columnwidth]{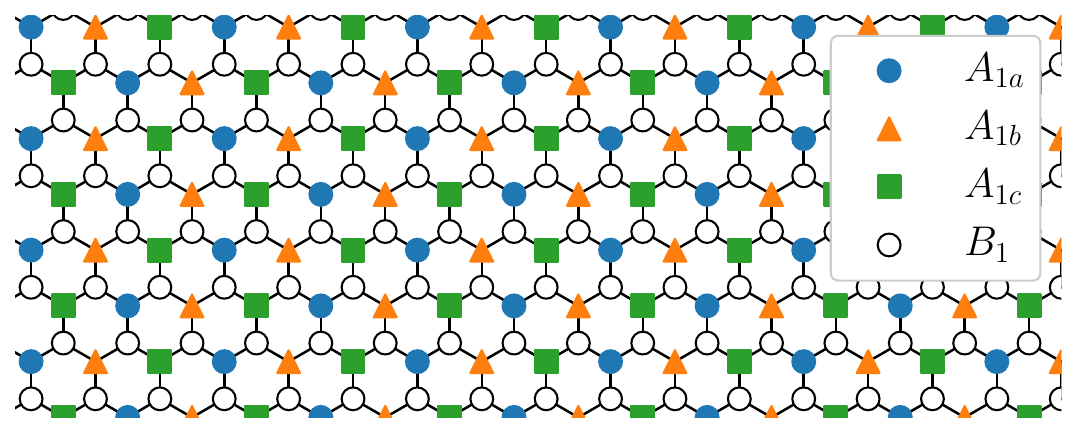}

  \caption{\raggedright
A diagram showing the sub-sublattice of one layer of BLG
  }
  \label{SI1}
\end{figure}
In the \textit{Eigenstates} section, we employ the sub-sublattice method to present
the real-space probability density distributions of selected eigenstates,
following the construction introduced in Ref.~\cite{PhysRevB.88.165405}.
This representation allows us to resolve the spatial structure of the eigenstates
in a manner compatible with the valley degrees of freedom of graphene.

The necessity of the sub-sublattice description arises from the fact that graphene
eigenstates generally involve contributions from both \(K\) and \(K^\prime\) valleys.
When such states are projected onto a single sublattice, the valley-dependent phase
factors lead to rapid oscillations of the wave-function amplitude at the atomic
scale. As a result, the probability density defined on the original Bravais lattice
exhibits strong site-to-site modulation, obscuring the physically relevant spatial
envelope of the eigenstates.

As illustrated in Fig.~\ref{SI1}, we now specify how the sub-sublattice
decomposition is implemented. We consider the atoms belonging to
sublattices \(A_1\) and \(B_1\) in the first graphene layer, and perform
the sub-sublattice construction on sublattice \(A_1\). Along the
horizontal direction, three consecutive \(A_1\) atoms are grouped into
a single unit, which defines three sub-sublattices denoted as
\(A_{1a}\), \(A_{1b}\), and \(A_{1c}\). These three sub-sublattices form
a reduced Bravais lattice with primitive vectors
\begin{equation}
\mathbf{a}_{1,\mathrm{sub}} = (3a,\,0), \qquad
\mathbf{a}_{2,\mathrm{sub}} = (\sqrt{3}a,\,0).
\end{equation}
With this construction, the probability density defined on each
sub-sublattice becomes continuous in real space.

To make explicit why this construction removes the atomic-scale
oscillations discussed above, we consider a single wave vector
\(\mathbf{k}\) contributing to the wave function on sublattice \(A_1\),
\begin{equation}
\bigl|\psi^{\mathbf{k}}\bigr\rangle_{A_1}
=
\frac{1}{\sqrt{N}}
\sum_{n,m}
\exp\!\left(i\,\mathbf{k}\cdot\mathbf{R}_{nm}\right)
\,|n,m\rangle .
\end{equation}
The position vector of an \(A_1\)-sublattice site is defined as
\begin{equation}
\mathbf{R}_{nm} \equiv |n,m\rangle
=
n\,\mathbf{a}_1 + m\,\mathbf{a}_2 ,
\end{equation}
where \(n\) and \(m\) are integers, and
\(\mathbf{a}_1=(a,0)\) and
\(\mathbf{a}_2=(\tfrac{1}{2}a,\tfrac{\sqrt{3}}{2}a)\) are the primitive
Bravais lattice vectors.

The two inequivalent valleys are located at
\begin{equation}
\mathbf{K} = \left(\frac{4\pi}{3a},\,0\right), \qquad
\mathbf{K}^\prime = \left(-\frac{4\pi}{3a},\,0\right).
\end{equation}
Substituting the valley wave vectors into the expression above yields
\begin{equation}
\bigl|\psi^{K(K^\prime)}\bigr\rangle_{A_1}
=
\frac{1}{\sqrt{N}}
\sum_{n,m}
\exp\!\left[
(-)\,\frac{2\pi i}{3}\,(2n+m)
\right]
\,|n,m\rangle .
\end{equation}

The phase factor
\(\exp\!\left[(-)\,2\pi i (2n+m)/3\right]\)
can take only three distinct values, which are identical for all sites
belonging to the same sub-sublattice, as illustrated in
Fig.~\ref{SI1}. Consequently, translational symmetry is restored within
each sub-sublattice provided that the primitive translation vectors are
chosen as
\(\mathbf{a}_{1,\mathrm{sub}} = (3a,\,0)\) and
\(\mathbf{a}_{2,\mathrm{sub}} = (\sqrt{3}a,\,0)\).

\section{Symmetry properties}
\begin{figure}[t]
  \centering
  \includegraphics[width=0.95\columnwidth]{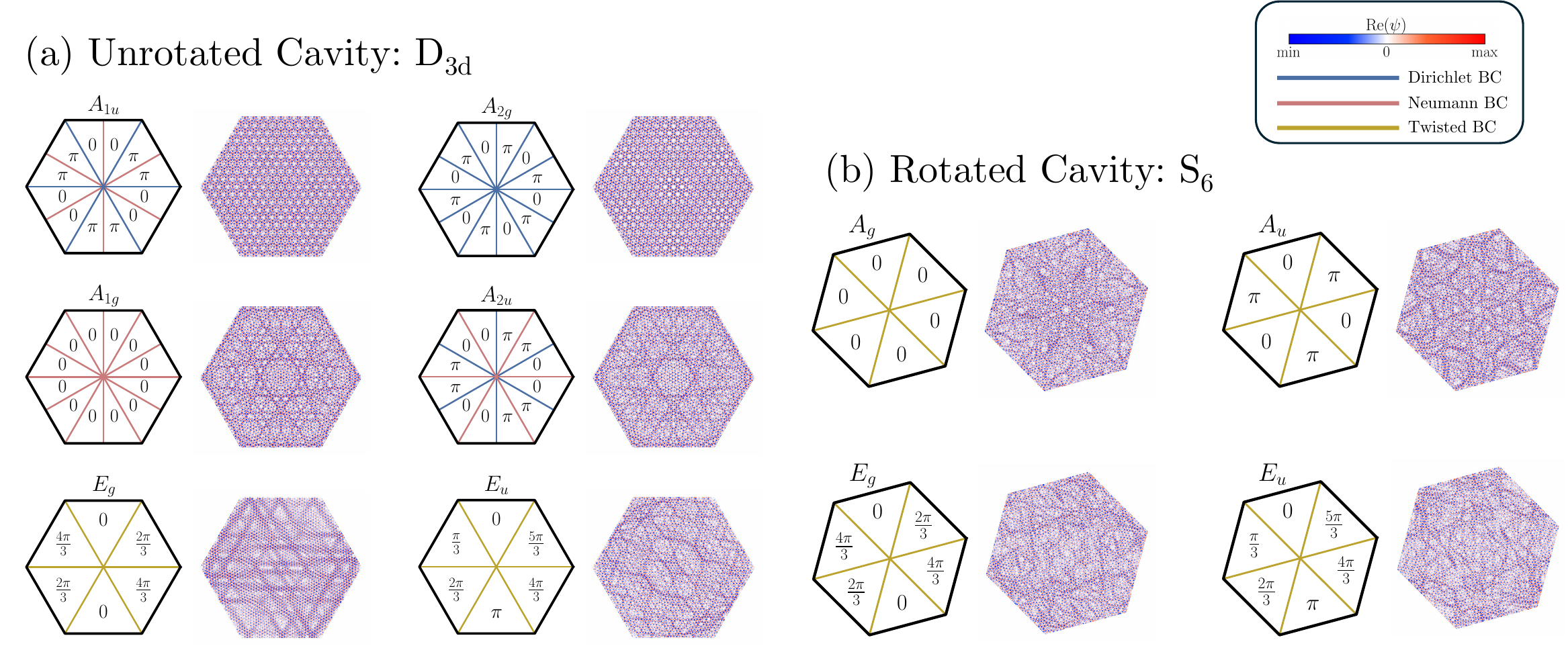}

  \caption{\raggedright
Symmetry properties of wavefunctions. (a) Unrotated cavity with $D_{3d}$ symmetry. The diagrams show the phase relations within each wedge and the corresponding boundary conditions for different subspaces of the $D_{3d}$ group, along with the real part of representative wavefunctions of eigenstates for each subspace. (b) Corresponding results for the rotated cavity with $S_6$ symmetry. For the $E$ subspaces of both unrotated and rotated cavities, we show the $m=1$ case. For the boundary lines of the wedges, the blue line denotes a Dirichlet boundary condition, where the wavefunction vanishes; the red line denotes a Neumann boundary condition, where the derivative of the wavefunction vanishes; and the yellow line denotes a twisted boundary condition, in which the wavefunction in the boundary is constrained with a phase relation aligned with the wedges.
  }
  \label{SI2}
\end{figure}

We illustrate the symmetry properties of the wavefunctions in both the unrotated and rotated cavities in Fig.~\ref{SI2}. Based on symmetry, the hexagon can be decomposed into wedges with specific boundary conditions, and the phase relations within each wedge are shown together with the corresponding boundary conditions. For the $A$ subspace of the unrotated cavity, the hexagon can be further decomposed into twelve right triangles. In the $A_{1u}$ and $A_{2g}$ subspaces, the diagonal lines of the hexagon satisfy Dirichlet boundary conditions, whereas in the $A_{1g}$ and $A_{2u}$ subspaces the diagonal lines satisfy Neumann boundary conditions. For the $E$ subspace of the unrotated cavity, the hexagon can be decomposed into six equilateral triangles. In this case, the wedges obey a phase relation and are subject to twisted boundary conditions, i.e., the wavefunction acquires a phase shift under a rotation operation, as illustrated in Fig.~\ref{SI2}(a). For the rotated cavity with $S_6$ symmetry, the hexagon can also be decomposed into six equilateral triangles with twisted boundary conditions, and the phase relations for the wedges are shown in Fig.~\ref{SI2}(b).

\section{Semiclassical ray-dynamics}
\subsection{Methods}
Our semiclassical ray construction is formulated in momentum space and is controlled by the
Fermi contour together with the associated group velocity.
Throughout this work we base the ray dynamics on the standard four-band continuum description of
AB-stacked bilayer graphene (~\eqref{fourband}). In particular, the trigonal-warping anisotropy that drives the nontrivial ray deflection originates
from the $\gamma_3$ term.

To implement the warped reflection law quantitatively at the energies studied here (e.g.,
$E_F\simeq 0.2~\mathrm{eV}$), we construct a numerical Fermi contour using the standard AB-stacked
tight-binding parameters ($t_0=3.16~\mathrm{eV}$, $t_1=0.381~\mathrm{eV}$, $t_3=0.38~\mathrm{eV}$,
$t_4=0.14~\mathrm{eV}$, and lattice constant $a=0.24595~\mathrm{nm}$), and evaluate the group velocity
$\mathbf{v}(\mathbf{k})=\nabla_{\mathbf{k}}E(\mathbf{k})$ along this contour.
In this regime the contour is a single closed, approximately convex curve.

For a wave packet confined to a single band $E(\mathbf{k})$ in a static scalar potential 
$V(\mathbf{r})$, the semiclassical equations of motion are 
\begin{equation}
\dot{\mathbf{r}}
= \nabla_{\mathbf{k}} E(\mathbf{k})
 - \dot{\mathbf{k}} \times \boldsymbol{\Omega}(\mathbf{k}),
\qquad
\hbar \dot{\mathbf{k}}
= - \nabla_{\mathbf{r}} V(\mathbf{r}),
\label{eq:semiclassical_simple}
\end{equation}
where $\boldsymbol{\Omega}(\mathbf{k})$ is the Berry curvature.  
Because BLG at low energy is gapless, 
$\boldsymbol{\Omega}(\mathbf{k}) = 0$, and no anomalous velocity term appears~\cite{PhysRevB.59.14915}.

Thus all semiclassical structure arises from the strongly anisotropic group velocity $\mathbf{v}(\mathbf{k}) = \nabla_{\mathbf{k}} E(\mathbf{k})$.

During free flight the band energy is conserved.  
When a ray encounters a hard wall with outward normal $\hat{\mathbf{n}}$, 
the reflected momentum $\mathbf{k}'$ must satisfy
\begin{equation}
E(\mathbf{k}')=E(\mathbf{k}),\qquad \mathbf{k}'\cdot\hat{\mathbf t}=\mathbf{k}\cdot\hat{\mathbf t}.
\label{eq:reflection_final}
\end{equation}
Equivalently, writing
$\mathbf{k} = k_t \hat{\mathbf{t}} + k_n \hat{\mathbf{n}}$
with $\hat{\mathbf{t}} = \hat{\mathbf{z}} \times \hat{\mathbf{n}}$,
the tangential component $k_t$ is preserved, 
while $k_n'$ is determined from energy conservation along with the flipped normal velocity~\cite{PhysRevB.107.205404}.

Implementing the semiclassical dynamics of a warped-dispersion billiard in a polygonal cavity
is numerically difficult. The regular hexagon is a pseudointegrable billiard whose dynamics is singular at the corners:
most of the boundary is flat, while the vertices act as isolated singular points.
In a direct ray-propagation scheme based on continuous time evolution and local boundary reflection,
trajectories can suffer from severe numerical instabilities, including \emph{grazing reflections} and
\emph{spurious trapping near corners}. Near a vertex, the result can depend on how the corner is handled.
Physically, exact vertex hits occur only for a measure-zero set of initial conditions;
for generic trajectories, collisions always occur on the interior of an edge, arbitrarily close to a corner if needed.
Resolving this behavior accurately requires very fine time steps and careful corner handling,
which becomes inefficient and unstable for long trajectories.

To overcome these difficulties, we rewrite the dynamics as a discrete mapping between successive boundary collisions. We tabulate the relation between the incidence angle $\varphi_{\mathrm{in}}$
and the outgoing angle $\varphi_{\mathrm{out}}$, determined by conservation of energy and tangential component $k_t$.
Each table contains about $1\times 10^5-3\times 10^5$ sampled angles.
This approach encodes the warped reflection law once and for all,
and removes the need to solve for $\varphi_{\mathrm{out}}$ by a local root search at every collision.

Between collisions, the next collision point is found by ray-boundary intersection.
The dynamics is thus an alternation of (i) intersection updates for free flight and (ii) table-based
updates for reflection.
This map-based scheme keeps high angular accuracy even for nearly grazing trajectories, and avoids artificial corner sticking and other corner-related numerical artifacts.
As a result, we can compute long trajectories and Poincar\'e sections robustly.

\begin{figure}[t!]
  \centering
  \begin{tikzpicture}
    \node[inner sep=0] (img) {\includegraphics[width=0.8\columnwidth]{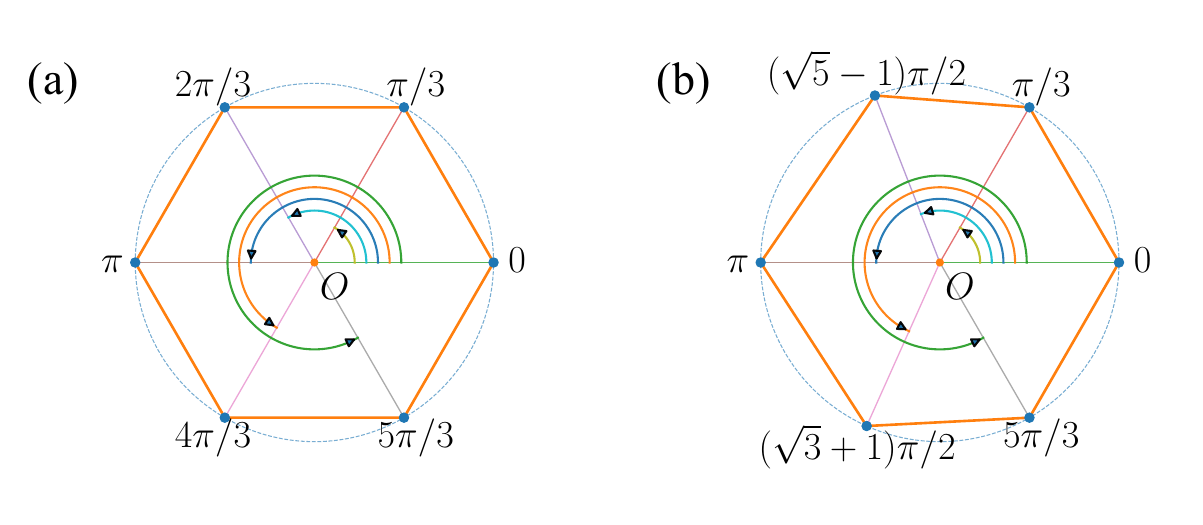}};
  \end{tikzpicture}

  \caption{\raggedright  Comparison of two hexagonal geometries inscribed in a circle of radius $R$. 
(a) a regular hexagon with equally spaced polar angles $0$, $\pi/3$, $2\pi/3$, $\pi$, $4\pi/3$, and $5\pi/3$. 
(b) an irrational hexagon with vertices at $0$, $\pi/3$, $(\sqrt{5}-1)\pi/2$, $\pi$, $(\sqrt{3}+1)\pi/2$, and $5\pi/3$. 
Central angles are measured counterclockwise from $\theta=0$ and indicated by staggered arcs with arrowheads.}

  \label{SI3}
\end{figure}

\subsection{Irrational hexagon billiard}

As a reference geometry, the regular hexagon in Fig.~\ref{SI3}(a) is obtained by placing six vertices uniformly on a circle of radius $R$ at the polar angles 
$0$, $\pi/3$, $2\pi/3$, $\pi$, $4\pi/3$, and $5\pi/3$. 
Successive vertices are connected by straight edges to form a closed polygonal cavity. 
This geometry retains discrete rotational symmetry and belongs to the class of pseudointegrable billiards. 

To induce genuinely ergodic semiclassical dynamics, we deliberately break this angular commensurability. The irrational hexagon in Fig.~\ref{SI3}(b) is constructed by modifying two of the six polar angles while keeping all vertices constrained to the same circumscribed circle of radius $R$. 
Specifically, the vertices are placed at $0$, $\pi/3$, $(\sqrt{5}-1)\pi/2$, $\pi$, $(\sqrt{3}+1)\pi/2$, $5\pi/3$,
measured counterclockwise from the reference direction $\theta=0$.

Geometrically, the cavity remains a convex polygon inscribed in a circle, but its edge orientations are now irrationally related. Dynamically, this suppresses the structured families of periodic orbits associated with commensurate angle relations and promotes a more uniform exploration of phase space, as evidenced by the rapid filling of the allowed region in the corresponding Poincaré sections.

\section{Further analysis of eigenstates}

\subsection{Angle dependence of correlation length}

\begin{figure}[t!]
  \centering
  \begin{tikzpicture}
    \node[inner sep=0] (img) {\includegraphics[width=0.9\columnwidth]{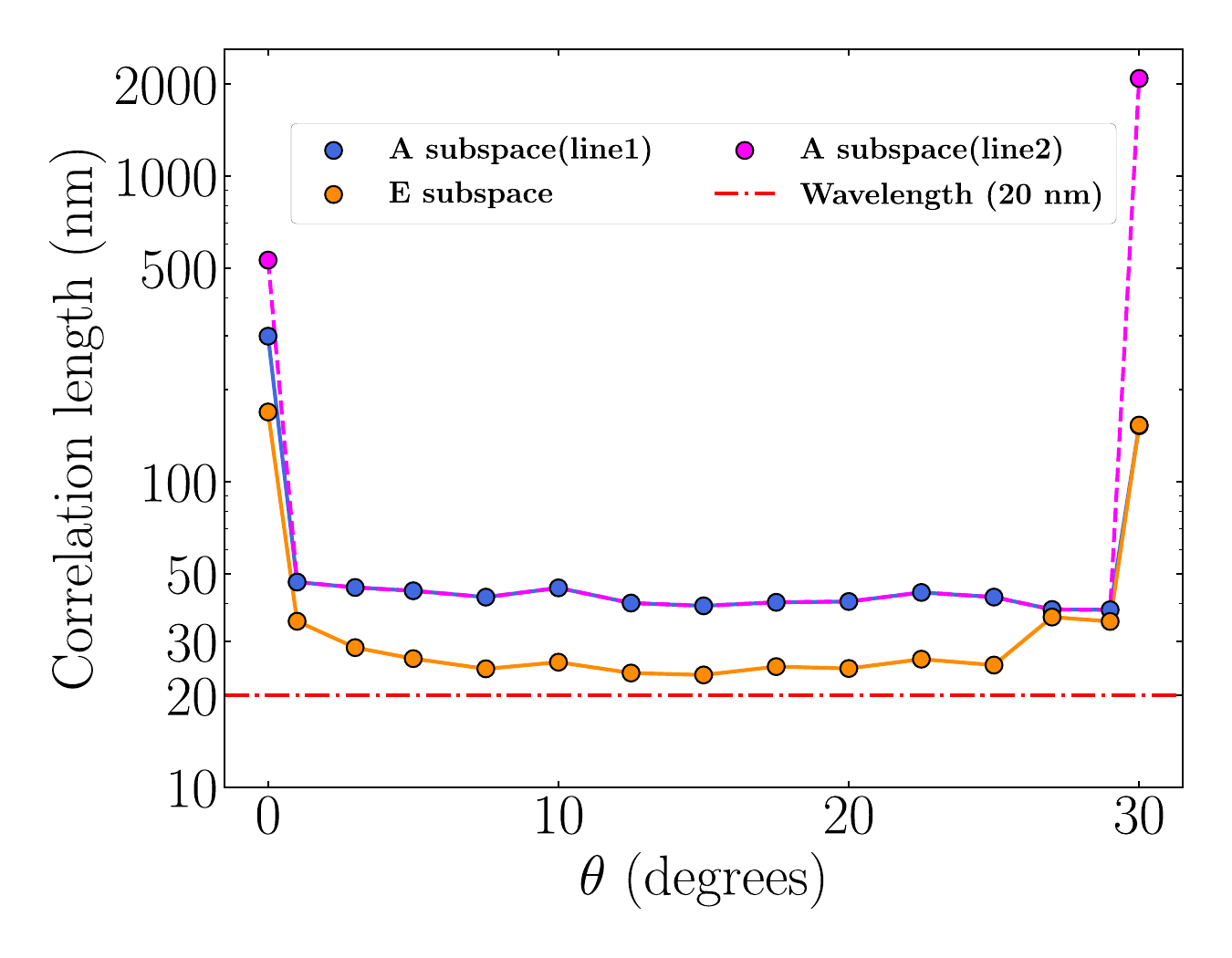}};
  \end{tikzpicture}

  \caption{\raggedright
Angle dependence of the correlation length on the rotation angle~\( \theta \) for the \( A \) and \( E \) subspaces. In the \( A \) subspaces, two curves are plotted. The blue solid curve corresponds to the pseudointegrable sector at \( \theta = 0^\circ \) and \( 30^\circ \) for the unrotated cavities, while the purple dashed curve denotes the integrable one.
}
  \label{SI4}
\end{figure}

Fig.~\ref{SI4} summarizes the dependence of the average correlation length on the rotation angle $\theta$ for different symmetry subspaces. In all cases, the correlation length rapidly decreases toward the wavelength scale upon rotation. This trend further supports the conclusion that misalignment between the lattice and the cavity boundary drives the system from a pseudointegrable regime to one exhibiting fully developed quantum-chaotic behavior.

\subsection{Momentum space IPR statistics}

\begin{figure}[t]
  \centering
  \includegraphics[width=0.95\columnwidth]{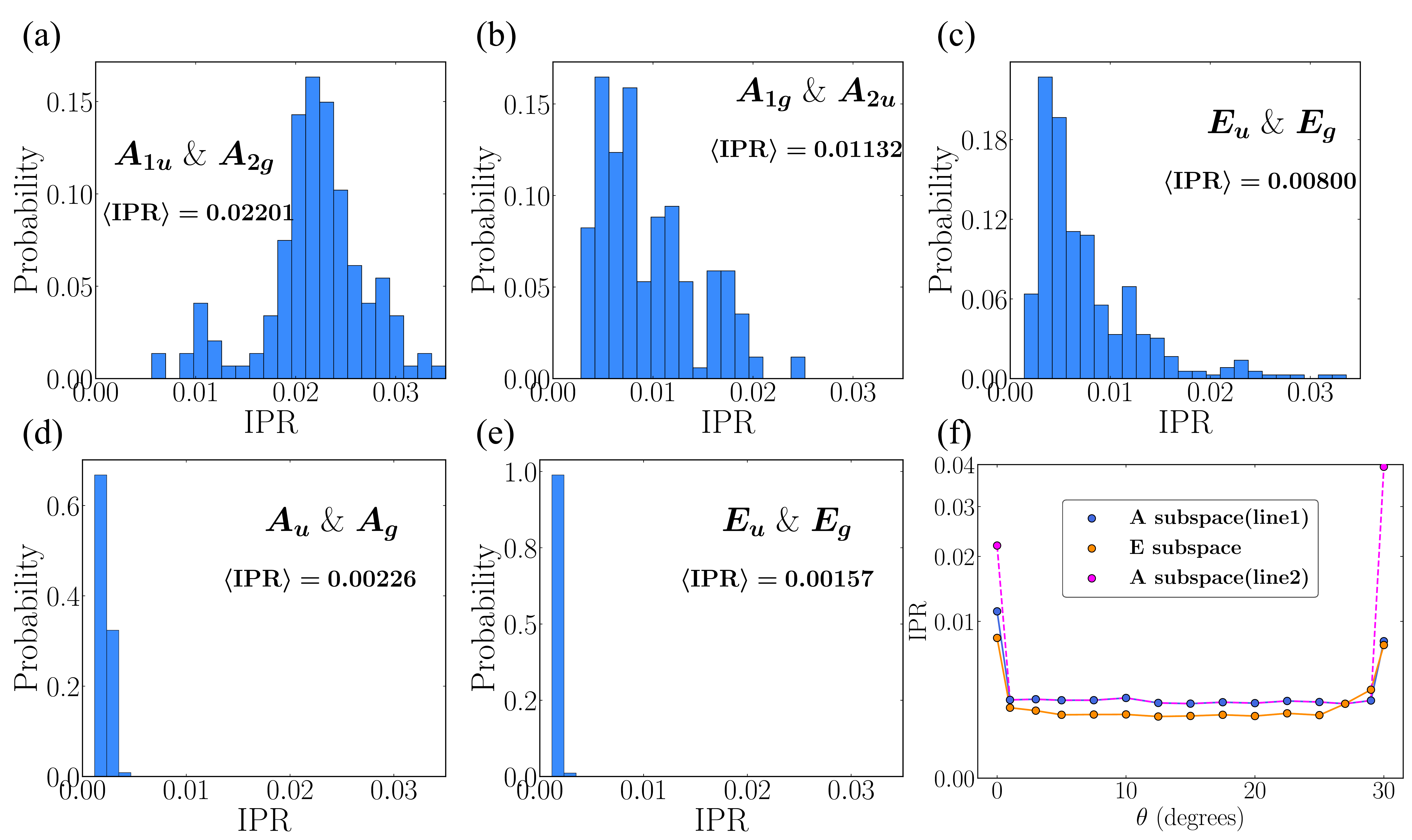}

  \caption{\raggedright
IPR of momentum space analysis for eigenstates. (a,b,c) Statistics of the IPR in momentum space for the eigenstates of unrotated cavities across different subspaces. (d,e) The corresponding results for rotated cavities. (f) The angle dependence of the IPR on the rotation angle $\theta$ for the A and E subspaces, where the second line of the A subspace starts and ends in the integrable sectors at $\theta=0^\circ$ and $\theta=30^\circ$.
  }
  \label{SI5}
\end{figure}

We further introduce an additional measure to illustrate how rotation drives the wavefunctions toward a Berry-wave-like structure. Specifically, we examine the number of plane waves composing each wavefunction, which can be quantified by the degree of localization in momentum space. Accordingly, we compute the inverse participation ratio (IPR) of the momentum-space density distribution $|\tilde{\psi}(\mathbf{k})|^2$, defined as

\[
\mathrm{IPR} = \frac{\sum_{\mathbf{k}} |\tilde{\psi}(\mathbf{k})|^4}{\left(\sum_{\mathbf{k}} |\tilde{\psi}(\mathbf{k})|^2 \right)^2}.
\]
A larger IPR indicates stronger localization in momentum space, corresponding to fewer plane-wave components contributing to the wavefunction.

For both unrotated and rotated cavities, we select a total of 1,056 eigenstates near $0.2\,\mathrm{eV}$ and perform statistical analysis of each symmetry subspace based on their momentum-space IPR. Fig.~\ref{SI5}(a)-(c) show the IPR distribution for eigenstates of the unrotated cavity, while Fig.~\ref{SI5}(d) and (e) present the corresponding results for rotated cavities with a rotation angle of $\theta=15^\circ$, with the average IPR indicated. We observe that the IPR decreases significantly, with only a small
variance in the IPR distribution upon rotation. Fig.~\ref{SI5}(f) summarizes the angle dependence of the average IPR, showing that the reduced IPR relative to the unrotated case is robust against changes in rotation angle. These results indicate that upon rotation, the wavefunctions consist of a larger number of plane wave components with different momenta, reflecting an increased degree of chaotic behavior.

\section{Results of more cases}
\subsection{Unrotated hexagon with kinks}
\begin{figure}[t]
  \centering
  \includegraphics[width=0.95\columnwidth]{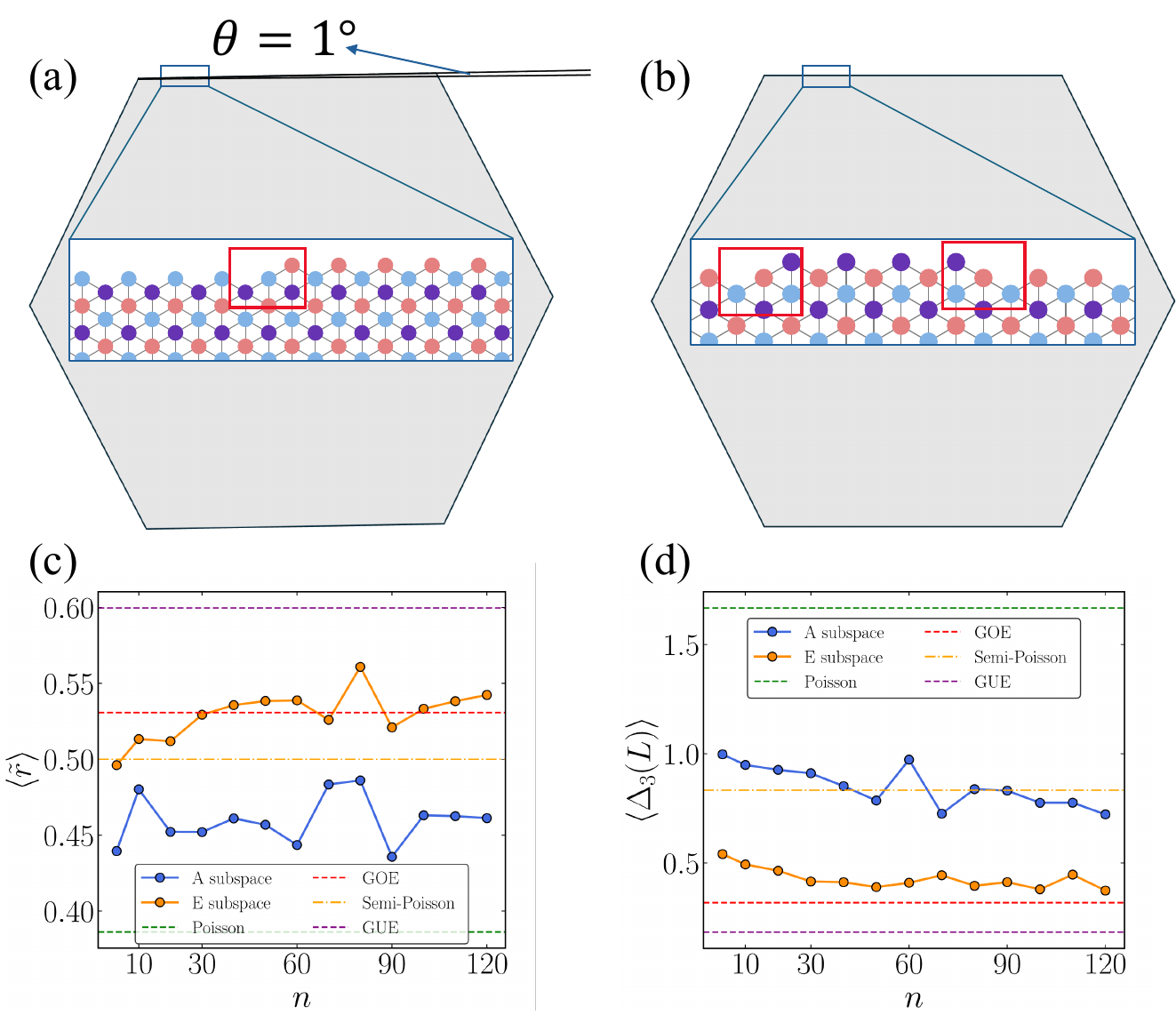}

  \caption{\raggedright
  Atomic geometry of the rotated cavities and unrotated cavities with kinks, the dependence of $\langle \tilde r \rangle$ and $\langle \Delta_3(L) \rangle$ on kinks number.
 (a) Rotated hexagon with a small rotation angle ($\theta = 1^\circ$); the inset shows the atomic profile at the cavity edge, with armchair edges highlighted in red and the remaining edges zigzag.  
(b) Unrotated hexagon with artificial edge disorder, red boxes indicate armchair edges.  
(c,d) Dependence of $\langle \tilde r \rangle$ (c) and $\langle \Delta_3(L) \rangle$ (d) on the number of added armchair kinks $n$, for the $S_6$ symmetry group.
  }
  \label{SI6}
\end{figure}

In addition to the ergodicity of the classical dynamics in the rotated hexagon, edge roughness constitutes another important ingredient that must be taken into account. To disentangle the roles of these two factors in producing quantum chaotic behavior, we consider a reference system in which kinks are introduced into an unrotated hexagon while preserving the $S_6$ symmetry. This system shares the same symmetry and comparable edge roughness with the rotated hexagon, but differs in its underlying classical dynamical correspondence. Consequently, the difference observed in their quantum chaotic behavior can be attributed to the trigonal-warping–induced modification of the classical dynamics.

In Fig.~\ref{SI6}(a), we show a hexagon rotated by an angle $\theta = 1^\circ$. The inset displays the atomic-scale edge profile, revealing detailed boundary structures. A mixture of zigzag and armchair terminations is observed; in particular, a representative turning point corresponding to an armchair edge is highlighted by a red box, which is a \textit{kink}. In Fig.~\ref{SI6}(b), we introduce the same type of armchair-edge kinks along the boundary of an unrotated hexagon while preserving the $S_6$ symmetry. 

We count the number of armchair-edge kinks and denote it by \(n\); for
reference, \(n = 11\) for \(\theta = 1^\circ\), and \(n\) follows an
approximately \(\sin\theta\)-dependent scaling. We then perform numerical simulations to investigate how the level spacing statistics evolve with increasing kink number. Fig.~\ref{SI6}(c) shows the behavior of $\langle \tilde r \rangle$. We find that the $A$ sectors remain at an intermediate value between the GOE and Poisson statistics, indicating the persistence of pseudointegrable behavior in this system. In contrast, the $E$ sectors exhibit an increasing $\langle \tilde r \rangle$ as more kinks are introduced. Nevertheless, even for large $n$, $\langle \tilde r \rangle$ does not reach the level of chaoticity observed in the rotated hexagon, where it remains robustly between the GOE and GUE values over a wide range of rotation angles. Fig.~\ref{SI6}(d) shows the behavior of the spectral rigidity $\langle \Delta_3(L) \rangle$. A similar trend is observed: the $A$ sectors remain at an intermediate value, while the $E$ sectors approach the GOE prediction instead of GUE prediction as in the rotated hexagon.

\subsection{Small angle case}

\begin{figure}[t]
  \centering
  \includegraphics[width=0.95\columnwidth]{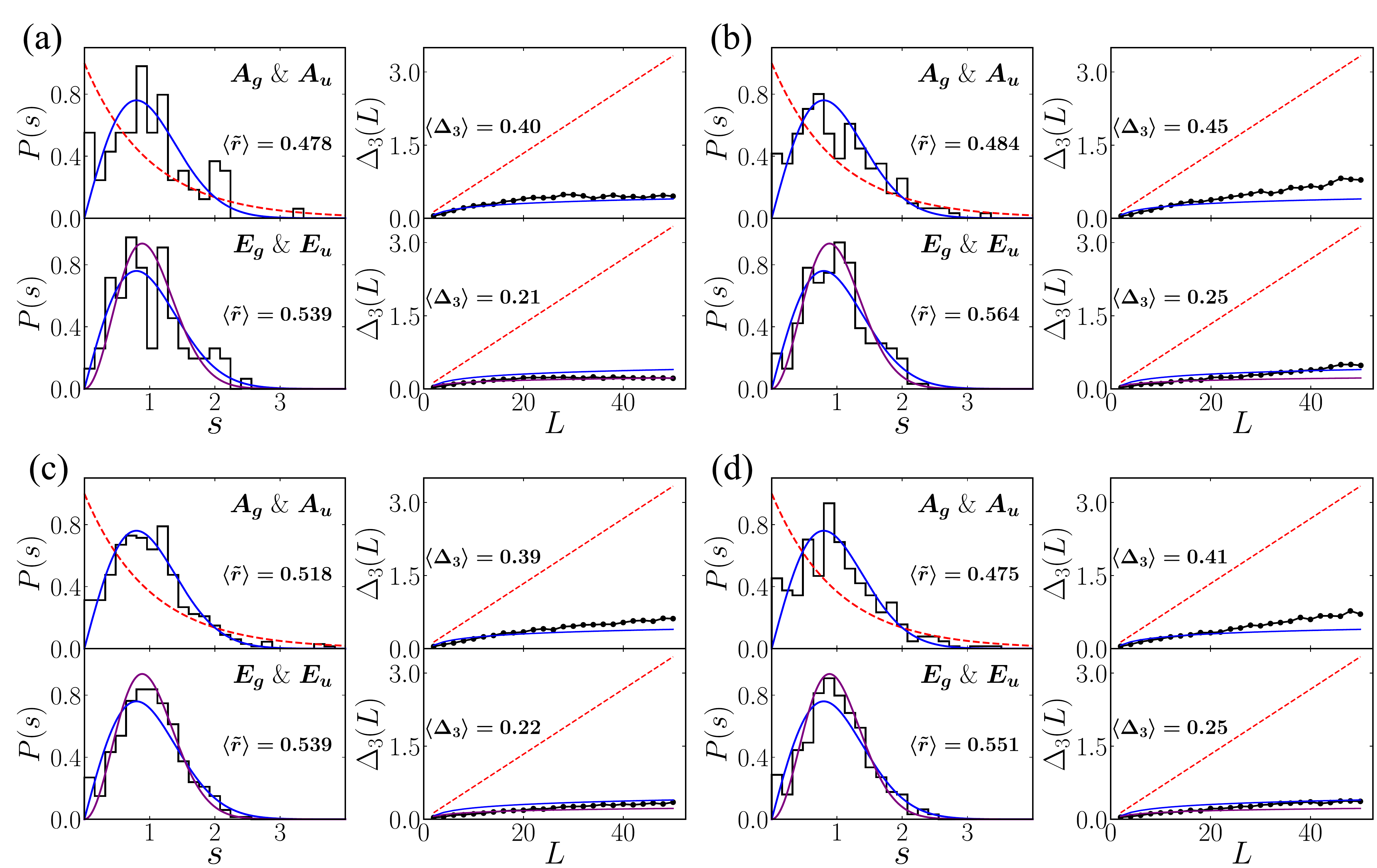}

  \caption{\raggedright
The figures illustrate the level spacing distributions and spectral rigidity for the small-angle cases, with the average $r$ value $\langle \tilde r \rangle$ and average rigidity $\langle \Delta_3(L) \rangle$ indicated. The results correspond to rotation angles $\theta$ of (a) $0.1^\circ$, (b) $0.2^\circ$, (c) $0.3^\circ$, and (d) $0.4^\circ$. The red dashed line denotes the Poisson distribution, the blue solid line represents the Wigner–Dyson distribution for the GOE, and the purple solid line represents the Wigner–Dyson distribution for the GUE.
  }
  \label{SI7}
\end{figure}

To illustrate the sensitivity of rotational effects on quantum chaotic behavior, we present four small angle cases with $\theta = 0.1^\circ, 0.2^\circ, 0.3^\circ,$ and $0.4^\circ$ in Fig.~\ref{SI7}, showing both the level spacing distribution and spectral rigidity for the $A$ and $E$ sectors. Even at these small angles, the $A$ subspaces exhibit a semi-Poisson-like level spacing distribution with $\langle \tilde r \rangle$ between $\langle \tilde r \rangle_{\mathrm{Poisson}}=0.386$ and $\langle \tilde r \rangle_{\mathrm{GOE}}=0.531$, whereas the $E$ subspaces display an intermediate distribution between GOE and GUE, with $\langle \tilde r \rangle$ between $\langle \tilde r \rangle_{\mathrm{GOE}}=0.531$ and $\langle \tilde r \rangle_{\mathrm{GUE}}=0.600$. In addition, the spectral rigidity exhibits signatures of chaotic behavior in both subspaces. For the $A$ subspaces, the rigidity closely follows the GOE prediction, whereas for the $E$ subspaces, it falls between the GOE and GUE predictions.

\subsection{Unsymmetrical case}

\begin{figure}[t]
  \centering
  \includegraphics[width=0.95\columnwidth]{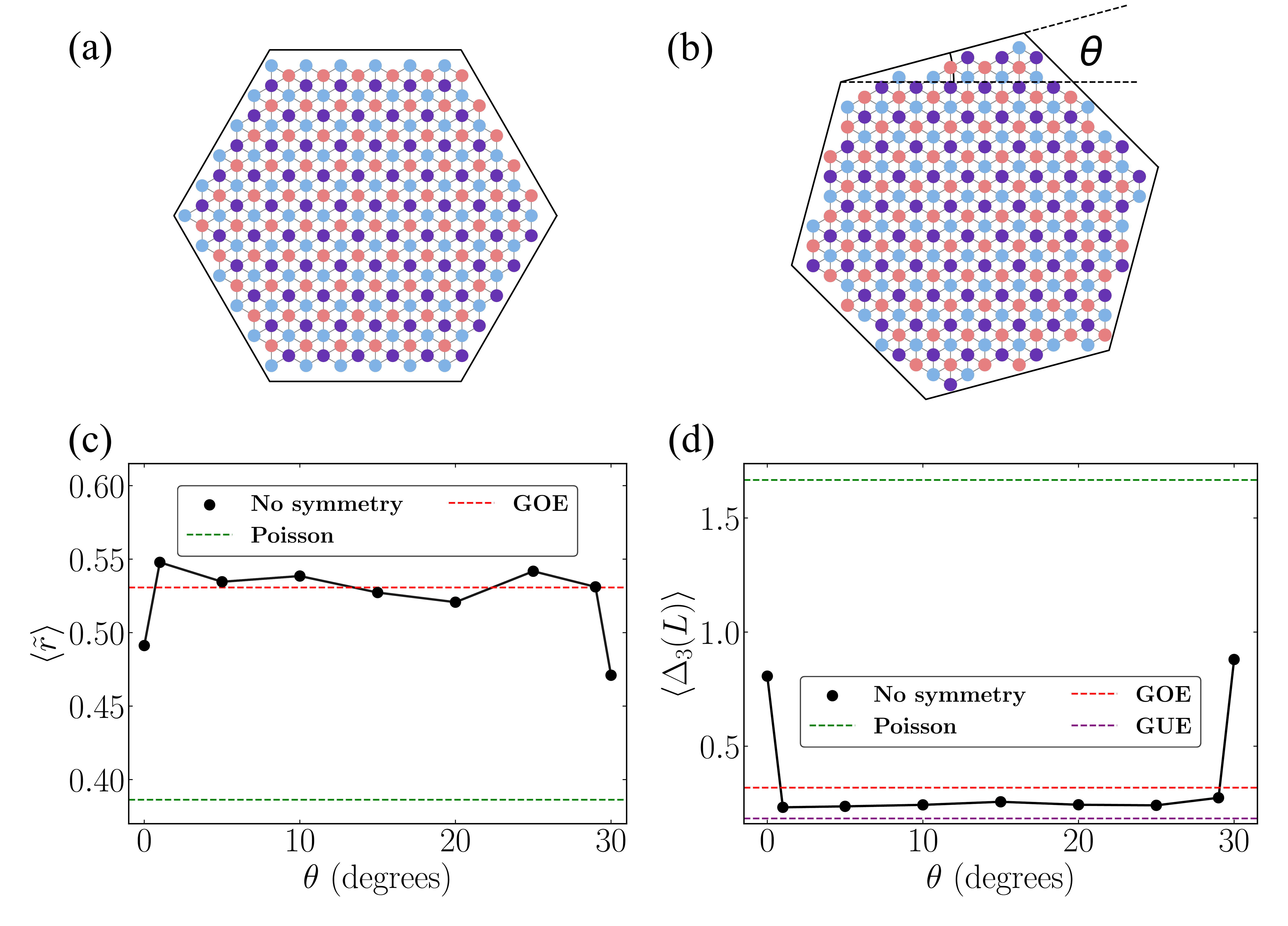}

  \caption{\raggedright
Atomic geometries of rotated and unrotated symmetry-broken BLG cavities, and the angle dependence of $\langle \tilde r \rangle$ and $\langle \Delta_3(L) \rangle$ on the rotation angle $\theta$.
(a) Atomic profile of an unrotated hexagon, exhibiting no symmetry.
(b) Corresponding profile for the rotated hexagon, which also lacks symmetry.
(c) Dependence of $\langle \tilde r \rangle$ on the rotation angle $\theta$.
(d) Dependence of $\langle \Delta_3(L) \rangle$ on the rotation angle $\theta$.
  }
  \label{SI8}
\end{figure}

We present another example of BLG cavities without point-group symmetry.
By shifting the hexagonal boundaries, the atomic point-group symmetry is
broken in both the unrotated and rotated cavities, as shown in
Fig.~\ref{SI8}(a) and (b). For this type of cavity, we perform the
eigenenergy analysis in the full Hilbert space without symmetry
decomposition.

The angle dependence of \(\langle \tilde r \rangle\) as a function of
the rotation angle \(\theta\), shown in Fig.~\ref{SI8}(c), reveals a
transition of \(\langle \tilde r \rangle\) from an intermediate value in
the unrotated cases at \(\theta = 0^\circ\) and \(\theta = 30^\circ\) to
a value close to the GOE prediction upon rotation. The intermediate
values of \(\langle \tilde r \rangle\) for the unrotated cavities reflect
the pseudointegrable nature of the hexagonal geometry, whereas the values of \(\langle \tilde r \rangle\) observed in the rotated cavities approaching GOE predication indicate the emergence of quantum chaotic behavior induced by rotation.

A similar trend is observed in the angle dependence of
\(\langle \Delta_3(L) \rangle\), as shown in Fig.~\ref{SI8}(d).
Specifically, \(\langle \Delta_3(L) \rangle\) remains at an intermediate
value for the unrotated cavities, while for the rotated cavities it
takes values between the GOE and GUE predictions.

\subsection{Other sizes}
\begin{figure}[t]
  \centering
  \includegraphics[width=0.95\columnwidth]{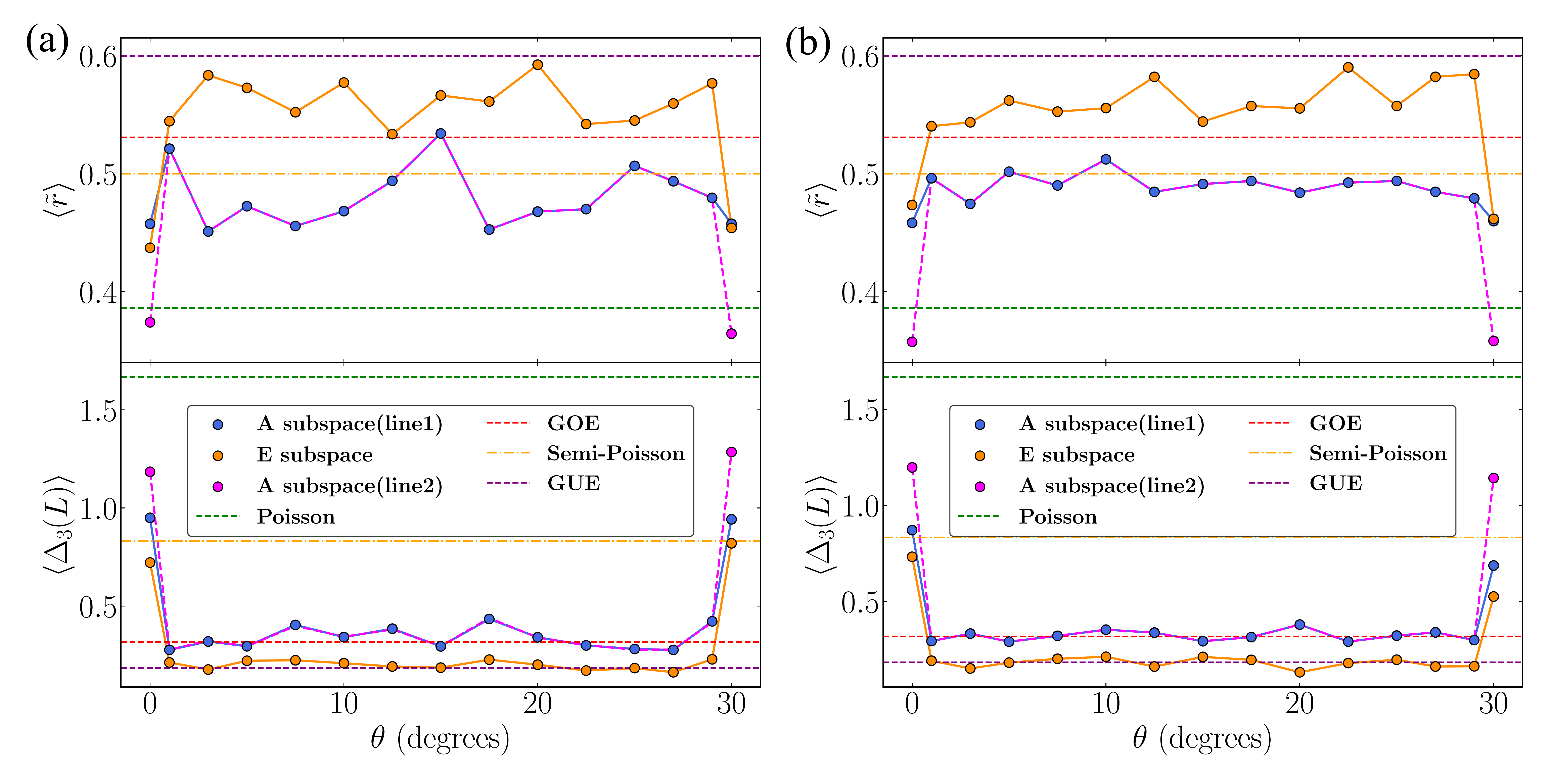}

  \caption{\raggedright
The figures illustrate the dependence of $\langle \tilde r \rangle$ and $\langle \Delta_3(L) \rangle$ on the rotation angle $\theta$ for cavities of different sizes: (a) $r = 450~\mathrm{nm}$, (b) $r = 500~\mathrm{nm}$.
  }
  \label{SI9}
\end{figure}

To rule out the possibility that the results presented in the main text arise from a special cavity size, we perform level statistics analyses for BLG cavities with other radii, $r = 450\,\mathrm{nm}$ and $r = 500\,\mathrm{nm}$. The angle dependence of $\langle \tilde r \rangle$ and $\langle \Delta_3(L) \rangle$ for different symmetry subspaces, shown in Fig.~\ref{SI9}, exhibits the same qualitative transition upon rotation.

\subsection{Monolayer graphene case}
\begin{figure}[t]
  \centering
  \includegraphics[width=0.95\columnwidth]{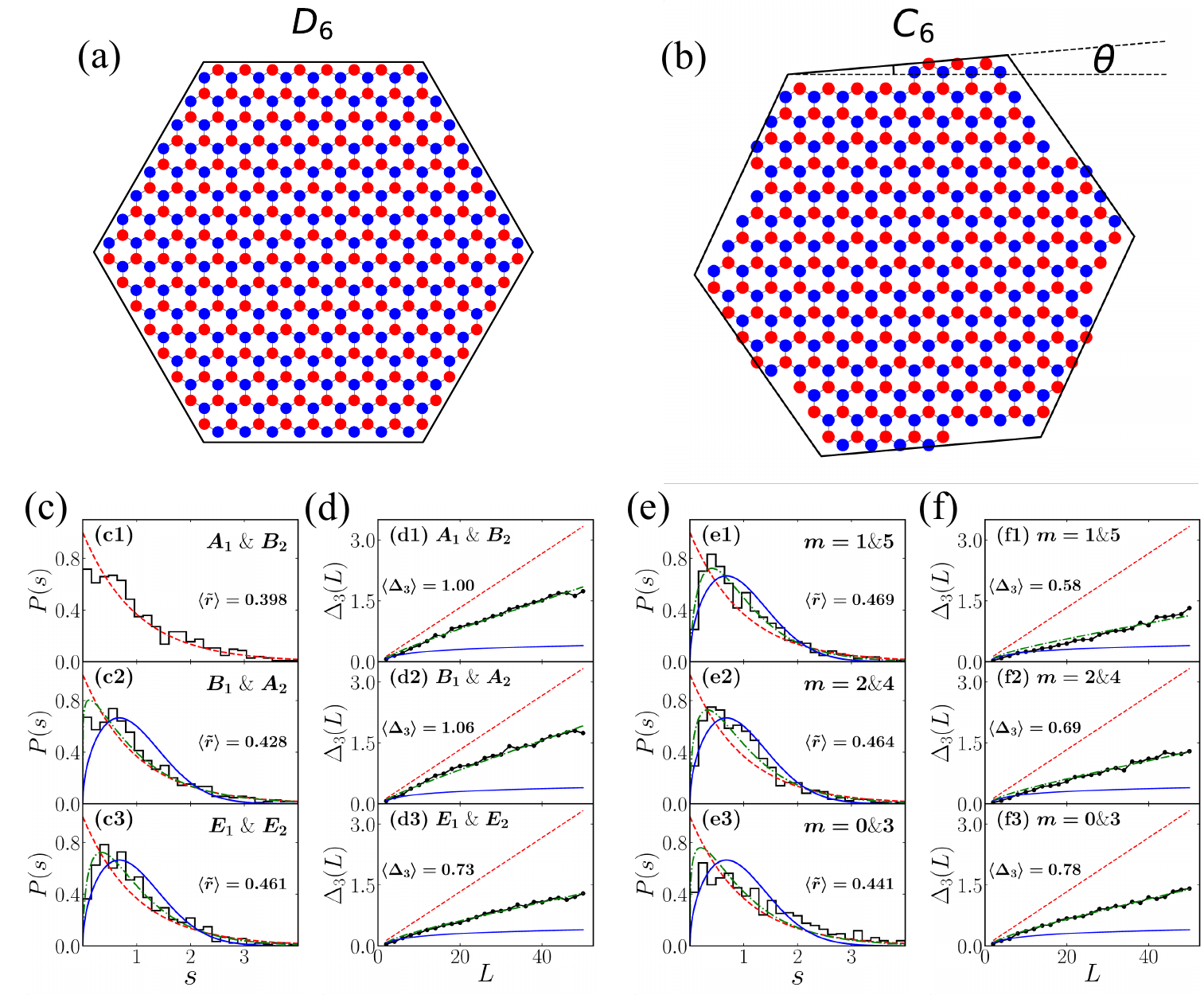}

  \caption{\raggedright
  Atomic geometries of rotated and unrotated MLG cavities and the corresponding level statistics results.
(a) Atomic configuration of an unrotated hexagonal MLG cavity exhibiting $D_6$ symmetry.
(b) Atomic configuration of a rotated hexagonal MLG cavity with a rotation angle $\theta = 5^\circ$, exhibiting $C_6$ symmetry.
(c,d) Nearest-neighbor level-spacing distribution and spectral rigidity for the unrotated MLG cavities.
(e,f) Corresponding results to panels (c) and (d) for the rotated MLG cavities with $\theta = 5^\circ$.
  }
  \label{SI10}
\end{figure}

In contrast to bilayer graphene cavities, we also investigate the quantum chaotic behavior of monolayer graphene (MLG) cavities. 

The unrotated MLG cavity preserves \(D_6\) symmetry, as shown in
Fig.~\ref{SI10}(a), while the rotated cavity preserves \(C_6\) symmetry,
as shown in Fig.~\ref{SI10}(b). The eigenstates of the cavity can be classified according to the
irreducible representations of the corresponding point group. The symmetry decomposition of the Hilbert space is determined by the rotational and mirror symmetries. 
The six-fold rotation operator \(C_6\) acts on the lattice sites and has eigenvalues
\begin{equation}
\lambda_m = e^{i 2\pi m / 6}, \qquad m = 0,1,2,3,4,5,
\end{equation}
which define the rotation quantum number \(m\) of each subspace. The mirror (reflection) operator \(\sigma\) acts along a symmetry axis of the hexagon and has eigenvalues \(\sigma = \pm 1\). Each irreducible subspace of the Hilbert space can thus be labeled by the pair \((m,\sigma)\).

For the unrotated cavity with \(D_6\) symmetry, both rotation and mirror symmetries are preserved, and the Hilbert space is decomposed into the subspaces \(A_1, A_2, B_1, B_2, E_1\) and \(E_2\), with the corresponding \(m\) and \(\sigma\) values indicated in Table~\ref{table}. For the rotated cavity with \(C_6\) symmetry, only the six-fold rotation is preserved, and the Hilbert space is decomposed into six one-dimensional subspaces \(\rho_0, \dots, \rho_5\), labeled by the rotation quantum number \(m=0,1,\dots,5\); mirror symmetry \(\sigma\) is no longer relevant. 
The full subspace labeling for both symmetries is summarized in Table~\ref{table}.

\begin{table}[t]
\centering
\caption{Subspace decomposition of the MLG cavity according to point group symmetries.}
\renewcommand{\arraystretch}{1.3}
\setlength{\tabcolsep}{10pt}

\begin{subtable}{0.45\textwidth}
\centering
\caption{$D_6$ symmetry}
\begin{tabular}{c|c|c|c}
\hline
\textbf{Irrep} & $\sigma$ & $m$ & $\text{Dim}$ \\
\hline
$A_1$ & $+1$ & $0$ & $1$ \\
$A_2$ & $-1$ & $0$ & $1$ \\
$B_1$ & $+1$ & $3$ & $1$ \\
$B_2$ & $-1$ & $3$ & $1$ \\
$E_1$ & $0$ & $1,5$ & $2$ \\
$E_2$ & $0$ & $2,4$ & $2$ \\
\hline
\end{tabular}
\end{subtable}
\hspace*{0.0\fill}
\begin{subtable}{0.45\textwidth}
\centering
\caption{$C_6$ symmetry}
\begin{tabular}{c|c|c}
\hline
\textbf{Irrep} & $m$ & $\text{Dim}$ \\
\hline
$\rho_0$ & $0$ & $1$ \\
$\rho_1$ & $1$ & $1$ \\
$\rho_2$ & $2$ & $1$ \\
$\rho_3$ & $3$ & $1$ \\
$\rho_4$ & $4$ & $1$ \\
$\rho_5$ & $5$ & $1$ \\
\hline
\end{tabular}
\end{subtable}
\label{table}
\end{table}

Based on the symmetry analysis of the MLG cavities, we perform level statistics calculations for the unrotated cavities, as shown in Fig.~\ref{SI10}(c) and (d). 
In Fig.~\ref{SI10}(c), the nearest-neighbor level spacing distributions for the \(A_1\) and \(B_2\) sectors follow the Poisson distribution, whereas the \(B_1\) and \(A_2\) sectors, as well as the two two-dimensional \(E\) sectors, exhibit semi-Poisson-like statistics. 
Correspondingly, the spectral rigidity for all subspaces, shown in Fig.~\ref{SI10}(d), is well described by the semi-Poisson-like fit.

For the rotated cavities, the same level statistics analysis is performed, as shown in Fig.~\ref{SI10}(e) and (f). Because of the approximate mirror symmetry of the rotated MLG cavity,
we remove the near-zero level spacings associated with the nearly degenerate
pairs in the $m=0$ and $m=3$ sectors.
Here, all subspaces display semi-Poisson-like statistics in the nearest-neighbor level spacing distributions (Fig.~\ref{SI10}(e)), and the corresponding spectral rigidity (Fig.~\ref{SI10}(f)) is also consistent with the semi-Poisson-like fit.

Further results show that rotation does not drive MLG cavities toward quantum chaos; the rotated MLG cavities still preserve the pseudointegrability of the classical limit and exhibit intermediate statistics close to semi-Poisson. This behavior arises from the isotropic Fermi surface of MLG in the low-energy Dirac regime, in contrast to the trigonal-warped Fermi surface in BLG cavities. These findings further justify the choice of BLG cavities as the platform for our study.

\bibliography{references_SI}